\documentclass[10pt]{article}
\usepackage{amssymb}
\usepackage{latexsym}
\usepackage{amsmath}
\usepackage{amscd}

\numberwithin{equation}{section}
\newcommand{\bR}{{\mathbb R}}

\newcommand{\bC}{{\mathbb C}}

\newcommand{\kC}{{\mathcal C}}
\newcommand{\kD}{{\mathcal D}}

\newcommand{\kH}{{\mathcal H}}

\newcommand{\kN}{{\mathcal N}}

\newcommand{\gotG}{{\mathfrak G}}
\newcommand{\gotH}{{\mathfrak H}}

\newcommand{\gotK}{{\mathfrak K}}

\newcommand{\gotL}{{\mathfrak L}}

\newcommand{\ga}{{\alpha}}

\newcommand{\gga}{{\gamma}}
\newcommand{\gG}{{\Gamma}}
\newcommand{\gk}{{\kappa}}

\newcommand{\gl}{{\lambda}}

\newcommand{\gO}{{\Omega}}
\newcommand{\gP}{{\Pi}}

\newcommand{\gS}{{\Sigma}}

\newcommand{\gt}{{\tau}}

\newcommand{\gT}{{\Theta}}

\newcommand{\gY}{{\Upsilon}}

\newcommand{\slim}{\,\mbox{\rm s-}\hspace{-2pt} \lim}

\newcommand{\real}{{\Re{\mathrm e\,}}}
\newcommand{\imag}{{\Im{\mathrm m\,}}}
\newcommand{\dom}{{\mathrm{dom\,}}}
\newcommand{\ran}{{\mathrm{ran\,}}}

\newcommand{\spann}{{\mathrm{sp}}}

\newcommand{\clo}{{\mathrm{clo}}}

\newcommand{\spa}{{\mathrm{span}}}
\newcommand{\clospa}{{\mathrm{clospan}}}

\newtheorem{thm}{Theorem}[section]
\newtheorem{prop}[thm]{Proposition}
\newtheorem{lem}[thm]{Lemma}
\newtheorem{cor}[thm]{Corollary}
\newtheorem{exam}[thm]{Example}
\newtheorem{defn}[thm]{Definition}
\newtheorem{rem}[thm]{Remark}

\newcommand{\ba}{\begin{array}}
\newcommand{\ea}{\end{array}}
\newcommand{\bea}{\begin{eqnarray}}
\newcommand{\eea}{\end{eqnarray}}
\newcommand{\bead}{\begin{eqnarray*}}
\newcommand{\eead}{\end{eqnarray*}}
\newcommand{\be}{\begin{equation}}
\newcommand{\ee}{\end{equation}}
\newcommand{\bed}{\begin{displaymath}}
\newcommand{\eed}{\end{displaymath}}
\newcommand{\bl}{\begin{lem}}
\newcommand{\el}{\end{lem}}
\newcommand{\bp}{\begin{prop}}
\newcommand{\ep}{\end{prop}}
\newcommand{\bt}{\begin{thm}}
\newcommand{\et}{\end{thm}}
\newcommand{\Label}{\label}
\newcommand{\bc}{\begin{cor}}
\newcommand{\ec}{\end{cor}}
\newcommand{\la}{\Label}

\newcommand{\br}{\begin{rem}}
\newcommand{\er}{\end{rem}}
\newcommand{\bd}{\begin{defn}}
\newcommand{\ed}{\end{defn}}

\newenvironment{proof}%
{\begin{sloppypar}\noindent{\bf Proof.}}%
{\hspace*{\fill}$\square$\end{sloppypar}\bigskip}

\def\sG{{\mathfrak G}}   \def\sH{{\mathfrak H}}   
   \def\sK{{\mathfrak K}}   \def\sL{{\mathfrak L}}

      \def\dC{{\mathbb C}}

      \def\dR{{\mathbb R}}

      \def\cC{{\mathcal C}}
\def\cD{{\mathcal D}}      
   \def\cH{{\mathcal H}}   
   \def\cK{{\mathcal K}}   
   \def\cN{{\mathcal N}}

\def\sp{{\text{\rm sp\,}}}
\def\mul{{\text{\rm mul\,}}}

\newcommand{\Imag}{\mbox{{\rm Im}}\,}
\newcommand{\Real}{\mbox{{\rm Re}}\,}

\title{\vspace{-1.0cm}Scattering Theory for Open Quantum Systems}

\author{
Jussi Behrndt\thanks{behrndt@math.tu-berlin.de}\\
Technische Universit\"{a}t Berlin\\
Institut f\"{u}r Mathematik\\
Stra\ss e des 17.\ Juni 136\\
D--10623 Berlin, Germany
\and
Mark M. Malamud\thanks{mmm@telnet.dn.ua}\\
Donetsk National University\\
Department of Mathematics\\
Universitetskaya 24\\
83055 Donetsk, Ukraine\\
\and
Hagen Neidhardt\thanks{neidhard@wias-berlin.de}\\
WIAS Berlin\\
Mohrenstr. 39\\
D-10117 Berlin, Germany\\
\\
Dedicated to Pavel Exner on the occasion of his $60^{th}$ birthday.
}

\date{\today}

\begin{document}

\maketitle

\vspace{-1.0cm}
\begin{abstract}
\noindent
Quantum systems which interact with their environment are often
mo\-deled by maximal dissipative operators or so-called Pseudo-Hamiltonians.
In this paper the scattering theory for such open systems is considered.
First it is assumed that a single maximal dissipative operator $A_D$
in a Hilbert space $\sH$ is used to describe an open quantum system. In this case the minimal self-adjoint
dilation $\widetilde K$ of $A_D$ can be regarded as the Hamiltonian of a closed system which contains
the open system $\{A_D,\sH\}$, but since $\widetilde K$ is necessarily not
semibounded from below, this model is difficult to interpret from a physical point of view.
In the second part of the paper an open quantum system is modeled with
a family $\{A(\mu)\}$ of maximal dissipative operators
depending on energy $\mu$, and it is shown that the open system can be embedded into
a closed system where the Hamiltonian is semibounded. Surprisingly it turns out that the
corresponding scattering matrix can be completely recovered from scattering matrices
of single Pseudo-Hamiltonians as in the first part of the paper.
The general results are applied to a class of Sturm-Liouville operators arising in
dissipative and quantum transmitting Schr\"{o}dinger-Poisson systems.
\end{abstract}



\section{Introduction}

Quantum systems which interact with their environment appear naturally in various physical problems and
have been intensively studied in the last decades, see e.g. the monographes \cite{BP,D,E}.
Such an open quantum system is often modeled with
the help of a maximal dissipative operator, i.e., a closed
linear operator $A_D$ in some Hilbert space $\gotH$ which satisfies
\begin{equation*}
\imag(A_Df,f) \le 0, \qquad f \in \dom(A_D),
\end{equation*}
and does not admit a proper extension in $\gotH$ with this property.
The dynamics in the open quantum system are described by the
contraction semigroup $e^{-itA_D}$, $t \ge 0$.
In the physical literature
the maximal dissipative operator $A_D$ is usually called a pseudo-Hamiltonian.
It is well known that $A_D$ admits a self-adjoint dilation $\widetilde K$
in a Hilbert space $\gotK$ which contains $\sH$ as a closed subspace, that is, $\widetilde K$
is a
self-adjoint operator in $\gotK$ and
\begin{equation*}
P_\gotH\bigl(\widetilde K-\lambda\bigr)^{-1}\upharpoonright_\gotH=(A_D-\lambda)^{-1}
\end{equation*}
holds for all $\lambda\in\bC_+ := \{z \in \bC: \imag(z) > 0\}$, cf. \cite{FN}. Since
the operator $\widetilde K$ is self-adjoint it can be regarded as the
Hamiltonian or so-called quasi-Hamiltonian
of a closed quantum system which contains the open quantum system $\{A_D,\sH\}$ as a subsystem.\\

In this paper we first assume that an open quantum system is described by a single
pseudo-Hamiltonian $A_D$ in $\sH$ and that $A_D$ is an extension
of a closed densely defined symmetric operator $A$ in $\sH$ with
finite equal deficiency indices. Then the
self-adjoint dilation $\widetilde K$ can be realized as a self-adjoint extension
of the symmetric operator $A\oplus G$ in $\gotK=\gotH\oplus L^2(\dR,\cH_D)$, where
$\cH_D$ is finite-dimensional and $G$ is the symmetric operator in $L^2(\dR,\cH_D)$ given by
\begin{equation*}
Gg:=-i\frac{d}{dx}\,g,\quad\dom(G)=\bigl\{g\in W^1_2(\dR,\cH_D): g(0)=0\bigr\},
\end{equation*}
see Section~\ref{sec3.1}. If $A_0$ is a self-adjoint extension of $A$ in $\gotH$
and $G_0$ denotes the usual self-adjoint momentum operator in $L^2(\dR,\cH_D)$,
\begin{equation*}
G_0g:=-i\frac{d}{dx}\,g,\quad\dom(G)=W^1_2(\dR,\cH_D),
\end{equation*}
then the dilation $\widetilde K$ can be regarded as a singular perturbation
(or more precisely a finite rank perturbation in resolvent sense) of
the ``unperturbed operator'' $K_0:= A_0 \oplus G_0$, cf. \cite{AK,K1}.
From a physical point of view $K_0$ describes a situation where both subsystems $\{A_0,\gotH\}$
and $\{G_0,L^2(\dR,\cH_D)\}$ do not interact
while $\widetilde K$ takes into account
an interaction of the subsystems.
Since the spectrum $\sigma(G_0)$ of the momentum operator
is the whole real axis, standard perturbation results
yield $\sigma(\widetilde K)=\sigma(K_0)=\dR$ and, in particular, $K_0$ and $\widetilde K$ are necessarily
not semibounded from below. For this reason $K_0$ and $\widetilde K$ are often called quasi-Hamiltonians rather
than Hamiltonians.

The pair $\{\widetilde K,K_0\}$ is a complete scattering system in $\gotK=\gotH\oplus L^2(\dR,\cH_D)$, that is,
the wave operators
\begin{equation*}
W_\pm(\widetilde K,K_0) := \slim_{t\to\pm\infty}e^{it\widetilde K}e^{-itK_0}P^{ac}(K_0)
\end{equation*}
exist and are complete, cf. \cite{AJS,BW,Wei1,Y}. Here $P^{ac}(K_0)$
denotes the orthogonal projection in $\gotK$ onto the absolutely
continuous subspace $\gotK^{ac}(K_0)$ of $K_0$. The scattering operator
\begin{equation*}
S(\widetilde K,K_0) := W_+(\widetilde K,K_0)^*W_-(\widetilde K,K_0)
\end{equation*}
of the scattering system $\{\widetilde K,K_0\}$ regarded as an operator in
$\gotK^{ac}(K_0)$ is unitary, commutes with the absolutely continuous part
$K_0^{ac}$ of $K_0$ and is unitarily equivalent to a multiplication operator
induced by a (matrix-valued) function $\{\widetilde S(\lambda)\}_{\lambda\in\dR}$
in a spectral representation $L^2(\dR,d\lambda,\cK_\lambda)$ of
$K_0^{ac}=A_0^{ac}\oplus G_0$, cf. \cite{BW}. The family $\{\widetilde S(\lambda)\}$ is called the
scattering matrix of the scattering system $\{\widetilde K,K_0\}$ and is one of the most important
quantities in the analysis of scattering processes.

In our setting the scattering matrix $\{\widetilde S(\lambda)\}$ decomposes into a $2\times 2$ block matrix function
in $L^2(\dR,d\lambda,\cK_\lambda)$ and it is
one of our main goals in Section~\ref{dilations} to show that the left upper corner
in this decomposition coincides with the scattering matrix $\{S_D(\lambda)\}$ of the dissipative
scattering system $\{A_D,A_0\}$, cf. \cite{N84,N86,N87}. The right lower corner of $\{\widetilde S(\lambda)\}$
can be interpreted as the Lax-Phillips scattering matrix $\{S^{LP}(\lambda)\}$ corresponding to the Lax-Phillips
scattering system $\{\widetilde K,\cD_-,\cD_+\}$. Here $\cD_\pm:=L^2(\dR_\pm,\cH_D)$ are
so-called
incoming and outgoing subspaces for the dilation $\widetilde K$,
we refer to \cite{BW,LP} for details on Lax-Phillips scattering theory.
The scattering matrices $\{\widetilde S(\lambda)\}$, $\{S_D(\lambda)\}$ and $\{S^{LP}(\lambda)\}$
are all explicitely expressed in terms of an "abstract" Titchmarsh-Weyl function $M(\cdot)$ and
a dissipative matrix $D$ which corresponds to the maximal dissipative operator $A_D$ in $\gotH$
and plays the role of an "abstract" boundary condition.
With the help of this representation of $\{S^{LP}(\lambda)\}$ we easily recover the famous relation
\begin{equation*}
S^{LP}(\lambda)=W_{A_D}(\lambda-i0)^*
\end{equation*}
found by Adamyan and Arov in \cite{AA1,AA2,AA3,AA4}
between the Lax-Phillips scattering matrix and the characteristic function $W_{A_D}(\cdot)$ of
the maximal dissipative operator $A_D$, cf. Corollary~\ref{adamyanarov}.
We point out that $M(\cdot)$ and $D$ are completely determined by the operators $A\subset A_0$
and $A_D$ from the inner system. This is interesting also from the viewpoint of inverse problems, namely,
the scattering matrix $\{\widetilde S(\lambda)\}$ of $\{\widetilde
K,K_0\}$, in particular, the Lax-Phillips scattering matrix $\{S^{LP}(\lambda)\}$
can be recovered having to disposal only the dissipative scattering system $\{A_D,A_0\}$, see
Theorem~\ref{dilscat} and Remark~\ref{invremdil}.

We emphasize that this simple and somehow straightforward embedding method
of an open quantum system into a closed quantum system by choosing a self-adjoint dilation
$\widetilde K$ of the pseudo-Hamiltonian $A_D$ is very convenient for mathematical
scattering theory, but difficult to legitimate from a physical point
of view, since the quasi-Hamiltonians $\widetilde K$ and $K_0$ are necessarily not semibounded from below.\\

In the second part of the paper we investigate open quantum systems which are
described by an appropriate chosen family of maximal dissipative operators $\{A(\mu)\}$,
$\mu\in\dC_+$, instead of a single
pseudo-Hamiltonian $A_D$. Similarly to the first part of the paper we assume that the maximal dissipative operators $A(\mu)$
are extensions of a fixed symmetric operator $A$ in $\gotH$ with equal finite deficiency indices.
Under suitable (rather weak) assumptions on the family $\{A(\mu)\}$ there exists a symmetric
operator $T$ in a Hilbert space $\gotG$ and a self-adjoint extension $\widetilde L$ of $L=A\oplus T$
in $\gotL=\gotH\oplus\gotG$ such that
\begin{equation}\label{qwe}
P_\gotH\bigl(\widetilde L-\mu\bigr)^{-1}\upharpoonright_\gotH=\bigl(A(\mu)-\mu\bigr)^{-1},
\qquad\mu\in\dC_+,
\end{equation}
holds, see Section~\ref{coupsection}. For example, in one-dimensional models for carrier transport
in semiconductors the operators $A(\mu)$ are regular Sturm-Liouville differential
operators in $L^2((a,b))$ with $\mu$-dependent dissipative boundary conditions and the "linearization"
$\widetilde L$ is a singular Sturm-Liouville operator in $L^2(\dR)$,
cf. \cite{klp,wr2,wr1,KL} and
Section~\ref{exam2}. We remark that one can regard and interpret relation \eqref{qwe} also
from an opposite point of view. Namely, if a self-adjoint operator $\widetilde L$ in a Hilbert space
$\gotL$ is given, then the compression
of the resolvent of $\widetilde L$ onto any closed subspace $\gotH$ of $\gotL$ defines a family of maximal dissipative
operators $\{A(\mu)\}$ via \eqref{qwe}, so that each closed quantum system $\{\widetilde L,\sL\}$
naturally contains open quantum subsystems $\{\{A(\mu)\},\sH\}$ of the type we investigate here.
Nevertheless, since from a purely mathematical point of view both approaches
are equivalent we will not explicitely discuss this second interpretation.

If $A_0$ and $T_0$ are self-adjoint extension of $A$ and $T$ in $\gotH$ and $\gotG$, respectively,
then again $\widetilde L$ can be regarded as a singular perturbation of the self-adjoint
operator $L_0:=A_0\oplus T_0$ in $\gotL$. As above $L_0$ describes a situation where the subsystems
$\{A_0,\gotH\}$ and $\{T_0,\gotG\}$ do not interact while $\widetilde L$ takes into
account a certain interaction. We note that if $A$ and $T$ have finite
deficiency indices, then
the operator $\widetilde L$ is semibounded from below if and only if $A$ and $T$ are semibounded from
below. Well-known results imply that the pair $\{\widetilde L,L_0\}$ is a complete scattering system
in the closed quantum system and
again the scattering matrix $\{\widetilde S(\lambda)\}$ decomposes into a $2\times 2$ block matrix function
which can be calculated in terms of abstract Titchmarsh-Weyl functions.

On the other hand it can be shown that the family $\{A(\mu)\}$, $\mu\in\dC_+$, admits a continuation to $\dR$,
that is,
the limit $A(\mu+i0)$ exists for a.e. $\mu\in\dR$ in the strong resolvent sense and
defines a maximal dissipative operator. The family $A(\mu+i0)$, $\mu\in\dR$,
can be regarded as a family of energy dependent pseudo-Hamiltonians in $\gotH$ and,
in particular, each pseudo-Hamiltonian $A(\mu+i0)$ gives rise to a quasi-Hamiltonian
$\widetilde K_\mu$ in $\gotH\oplus L^2(\dR,\cH_\mu)$, a complete scattering system
$\{\widetilde K_\mu,A_0\oplus -i\tfrac{d}{dx}\}$ and a corresponding scattering matrix $\{\widetilde S_\mu(\lambda)\}$
as illustrated in the first part of the introduction.

One of our main observations in Section~\ref{couplescat} is that the scattering matrix
$\{\widetilde S(\lambda)\}$ of the scattering system $\{\widetilde L,L_0\}$ in $\gotH\oplus\gotG$
is related  to the scattering matrices $\{\widetilde S_\mu(\lambda)\}$ of the systems
$\{\widetilde K_\mu,A_0\oplus -i\tfrac{d}{dx}\}$, $\mu\in\dR$, in $\gotH\oplus L^2(\dR,\cH_\mu)$ via
\begin{equation}\label{mainres}
\widetilde S(\mu)=\widetilde S_\mu(\mu)\quad\text{for a.e.}\,\,\,\,\mu\in\dR.
\end{equation}
In other words, the scattering matrix $\{\widetilde S(\lambda)\}$ of
the scattering system $\{\widetilde L,L_0\}$
can be completely recovered from scattering matrices of scattering systems
for single quasi-Hamiltonians. Furthermore, under certain continuity properties of
the abstract Titchmarsh Weyl functions this implies $\widetilde S(\lambda)\approx \widetilde S_\mu(\lambda)$
for all $\lambda$ in a sufficiently small neighborhood of the fixed
energy $\mu\in\dR$, which legitimizes the concept of single quasi-Hamiltonians
for small energy ranges.

Similarly to the case of a single pseudo-Hamiltonian the diagonal entries of
$\{\widetilde S(\mu)\}$ or $\{\widetilde S_\mu(\mu)\}$ can be interpreted as scattering matrices
corresponding to energy dependent dissipative scattering systems and energy-dependent Lax-Phillips
scattering systems. Moreover, if $\{S^{LP}_\mu(\lambda)\}$ is the scattering matrix
of the Lax-Phillips scattering system $\{\widetilde K_\mu,L^2(\dR_\pm,\cH_\mu)\}$ and
$W_{A(\mu)}(\cdot)$ denote the characteristic functions of the maximal dissipative
operators $A(\mu)$ then an energy-dependent modification
\begin{equation*}
S^{LP}_\mu(\mu)=W_{A(\mu)}(\mu-i0)^*
\end{equation*}
of the classical Adamyan-Arov result holds for a.e. $\mu\in\dR$, cf. Section~\ref{mainsec}.\\

The paper is organized as follows. In Section~\ref{two} we give a brief
introduction into extension and spectral theory of symmetric and self-adjoint operators
with the help of boundary triplets and associated Weyl functions.
These concepts will play an important role throughout the paper.
Furthermore, we recall a recent result on the representation
of the scattering matrix of a scattering system consisting of two
self-adjoint extensions of a symmetric operator from \cite{BMN}.
Section~\ref{dilations} is devoted to open quantum systems described by a single
pseudo-Hamiltonian $A_D$ in $\gotH$.
In Theorem~\ref{III.1} a minimal
self-adjoint dilation $\widetilde K$ in $\gotH\oplus L^2(\dR,\cH_D)$ of the maximal dissipative operator $A_D$
is explicitely constructed. Section~\ref{dilatscat} and Section~\ref{laxsubsec} deal with the scattering matrix
of $\{\widetilde K,K_0\}$ and the interpretation of the diagonal entries as scattering
matrices of the dissipative scattering system $\{A_D,A_0\}$ and the Lax-Phillips scattering system
$\{\widetilde K,L^2(\dR_\pm,\cH_D)\}$. In Section~\ref{III} we give an example of a pseudo-Hamiltonian
which arises in the theory of dissipative Schr\"odinger-Poisson
systems, cf. \cite{BKNR1,BKNR2,KNR1}.
In Section~\ref{couplescat} the family $\{A(\mu)\}$ of maximal dissipative operators in $\gotH$
is introduced and, following ideas of \cite{DHMS00}, we construct a self-adjoint operator
$\widetilde L$ in a Hilbert space $\gotL$, $\gotH\subset\gotL$, such that \eqref{qwe} holds.
After some preparatory work the relation \eqref{mainres} between the scattering matrices of $\{\widetilde L,L_0\}$ and
the scattering systems consisting of quasi-Hamiltonians is verified in Section~\ref{mainsec}.
Finally, in Section~\ref{exam2}
we consider a so-called quantum transmitting Schr\"{o}dinger-Poisson system
as an example for an open quantum system which consists of a family of energy-dependent
pseudo-Hamiltonians, cf. \cite{klp,bdm,BF,wr2,wr1,KL}.\\

\noindent
{\bf Acknowledgment.} The authors thank Professor Peter Lax for helpful comments and fruitful
discussions. Jussi Behrndt gratefully acknowledges support by DFG, Grant 3765/1;
Hagen Neidhardt gratefully acknowledges support by DFG, Grant 1480/2.\\

\noindent
{\bf Notations.}  Throughout this paper $(\gotH,(\cdot,\cdot))$ and
$(\gotG,(\cdot,\cdot))$ denote separable Hilbert spaces. The linear space
of bounded linear
operators defined on ${\gotH}$ with values in ${\gotG}$ will be denoted by
$[{\gotH}, {\gotG}]$. If $\gotH={\gotG}$ we simply write $[{\gotH}]$. The set of
closed operators in $\gotH$ is denoted by $\kC(\gotH)$. The
resolvent set $\rho(S)$ of a closed linear operator $S\in\kC(\gotH)$ is the set of all
$\gl\in\dC$ such that $(S-\gl)^{-1}\in [\gotH]$, the spectrum $\sigma(S)$ of $S$
is the complement of $\rho(S)$ in $\dC$. $\sigma_p(S)$, $\sigma_c(S)$, $\sigma_{ac}(S)$ and $\sigma_r(S)$
stand for the point, continuous, absolutely continuous and residual spectrum of $S$, respectively.
The domain,
kernel and range of a linear operator
are denoted by $\dom(\cdot)$, $\ker(\cdot)$ and $\ran(\cdot)$, respectively.

\section{Self-adjoint extensions and scattering systems}\label{two}

In this section we briefly review the notion of abstract boundary triplets
and associated Weyl functions in the extension theory of symmetric operators,
see e.g. \cite{DM87,DM91,DM95,GG}.
For scattering systems consisting of a pair of self-adjoint extensions of
a symmetric operator with finite deficiency indices we recall a result on the
representation of the scattering matrix in terms of a Weyl function proved in \cite{BMN}.

\subsection{Boundary triplets and closed extensions}\label{btrips}

Let $A$ be a densely defined closed symmetric operator in
the separable Hilbert space
$\gotH$ with equal deficiency indices $n_\pm(A)=\dim\ker(A^*\mp i)\leq\infty$.
We use the concept of boundary
triplets for the description of the closed extensions
$A_\Theta\subseteq A^*$ of $A$ in $\gotH$.

\begin{defn}
A triplet $\Pi=\{\kH,\gG_0,\gG_1\}$ is called a {\rm boundary triplet} for the adjoint
operator $A^*$ if $\kH$ is a Hilbert space and
$\Gamma_0,\Gamma_1:\  \dom(A^*)\rightarrow\kH$ are linear mappings such that
the "abstract Green identity"
\begin{equation*}
(A^*f,g) - (f,A^*g) = (\gG_1f,\gG_0g) - (\gG_0f,\gG_1g),
\end{equation*}
holds for all $f,g\in\dom(A^*)$ and the map
$\gG:=(\Gamma_0,\Gamma_1)^\top:  \dom(A^*) \rightarrow \kH
\times \kH$ is surjective.
\end{defn}

We refer to \cite{DM91} and \cite{DM95} for a detailed study of
boundary triplets and recall only some important facts. First of all
a boundary triplet $\Pi=\{\kH,\gG_0,\gG_1\}$ for $A^*$ exists since
the deficiency indices $n_\pm(A)$ of $A$ are assumed to be equal.
Then $n_\pm(A) = \dim\kH$ and $A=A^*\upharpoonright\ker(\Gamma_0)\cap\ker(\Gamma_1)$ holds.
We note that a boundary triplet for
$A^*$ is not unique.

In order to describe the
closed extensions $A_\Theta\subseteq A^*$ of $A$ with
the help of a boundary triplet $\Pi=\{\kH,\Gamma_0,\Gamma_1\}$ for
$A^*$ we have to consider the set $\widetilde\kC(\kH)$ of closed
linear relations in $\kH$, that is, the set of closed linear
subspaces of $\kH\times\kH$. We usually use a column vector notation for the elements in a
linear relation $\Theta$. A closed linear operator in $\kH$ is
identified with its graph, so that the set  $\kC(\kH)$  of closed
linear operators in $\kH$ is viewed as a subset of
$\widetilde\kC(\kH)$, in particular, a linear relation $\Theta$ is an operator if and only if the
multivalued part
$\mul \Theta=\bigl\{ f^\prime:\bigl(\begin{smallmatrix} 0 \\ f'\end{smallmatrix}\bigr)\in\Theta\bigr\} $
is trivial.
For the usual definitions of the linear
operations with linear relations, the inverse, the resolvent set and
the spectrum we refer to \cite{DS87}.
Recall that the adjoint
relation $\Theta^*\in\widetilde\kC(\kH)$ of a linear relation
$\Theta$ in $\kH$ is defined as
\begin{equation*}
\Theta^*= \left\{
\begin{pmatrix} k\\k^\prime
\end{pmatrix}: (h^\prime,k)=(h,k^\prime)\,\,\text{for all}\,
\begin{pmatrix} h\\h^\prime\end{pmatrix}
\in\Theta\right\}
\end{equation*}
and $\Theta$ is said to be {\it symmetric} ({\it self-adjoint}) if
$\Theta\subset\Theta^*$ (resp. $\Theta=\Theta^*$). Notice that this
definition extends the definition of the adjoint operator.
For a self-adjoint relation $\Theta=\Theta^*$ in $\cH$ the multivalued part
$\mul \Theta$
is the orthogonal complement of $\dom \Theta$ in $\cH$. Setting
$\cH_{\rm op}:=\overline{\dom \Theta}$ and $\cH_\infty=\mul \Theta$ one verifies
that $\Theta$ can be written as the direct orthogonal sum of a self-adjoint
operator $\Theta_{\rm op}$ in the Hilbert space $\cH_{\rm op}$ and the ``pure''
relation $\Theta_\infty=\bigl\{\bigl(\begin{smallmatrix} 0 \\ f'
\end{smallmatrix}\bigr):f'\in\mul \Theta\bigr\}$ in the Hilbert
space $\cH_\infty$.

A linear
relation $\Theta$ in $\kH$ is called {\it dissipative} if
$\imag(h',h)\leq 0$ holds for all $(h,h')^\top\in \Theta$
and $\Theta$ is called {\it maximal
dissipative} if it is dissipative and does not admit proper dissipative
extensions in $\cH$; then $\Theta$ is necessarily closed, $\Theta\in\widetilde\cC(\cH)$.
We remark that a linear relation $\Theta$ is maximal
dissipative if and only if $\Theta$ is dissipative and some $\lambda\in\bC_+$
(and hence every $\lambda\in\bC_+$) belongs to $\rho(\Theta)$.

A description of all closed (symmetric, self-adjoint, (maximal)
dissipative) extensions of $A$ is given in the next proposition.

\begin{prop}\label{propo}
Let $A$ be a densely defined closed symmetric operator in $\gotH$ with equal deficiency indices and
let
$\Pi=\{\kH,\gG_0,\gG_1\}$ be a boundary triplet for  $A^*.$  Then the mapping
\begin{equation}\label{bij}
\Theta\mapsto A_\Theta:= A^*\upharpoonright \Gamma^{(-1)}\Theta=
A^*\upharpoonright \bigl\{f\in\dom(A^*): \  (\Gamma_0
f,\Gamma_1 f)^\top\in\Theta\bigr\}
\end{equation}
establishes  a bijective correspondence between the set
$\widetilde\kC(\kH)$ and the set of closed extensions $A_\Theta\subseteq A^*$ of $A$.
Furthermore
\begin{equation*}
(A_\Theta)^*=  A_{\Theta^*}
\end{equation*}
holds for any $\Theta\in\widetilde\kC(\kH)$. The extension $A_\Theta$ in \eqref{bij}
is symmetric (self-adjoint, dissipative,
maximal dissipative) if and only if $\Theta$ is symmetric (self-adjoint, dissipative,
maximal dissipative).
\end{prop}

It follows immediately from this proposition that if $\Pi=\{\kH,\gG_0,\gG_1\}$
is a boundary triplet for $A^*$, then the extensions
\begin{equation*}
A_0:=A^*\!\upharpoonright\ker(\gG_0)
\quad \text{and}\quad
A_1:=A^*\!\upharpoonright\ker(\gG_1)
\end{equation*}
are self-adjoint. In the sequel usually the extension $A_0$ corresponding to the boundary mapping
$\Gamma_0$ is regarded as a "fixed" self-adjoint extension.
We note that the closed extension $A_\Theta$ in
\eqref{bij} is disjoint with $A_0$, that is $\dom(A_\Theta)\cap \dom(A_0) =\dom(A),$ if
and only if $\Theta\in \kC(\kH)$. In this case
\eqref{bij} takes the form
\begin{equation}\label{bijop}
A_\Theta=A^*\!\upharpoonright
\ker\bigl(\Gamma_1-\Theta\Gamma_0\bigr).
\end{equation}

For simplicity we will often restrict ourselves to simple symmetric
operators. Recall that a symmetric operator is said to be {\it
simple} if there is no nontrivial subspace which reduces it to a
self-adjoint operator. By \cite{K49} each symmetric operator $A$ in
$\gotH$ can be written as the direct orthogonal sum $\widehat A\oplus
A_s$ of a simple symmetric operator $\widehat A$ in the Hilbert
space
\begin{equation*}
\widehat\gotH=\clo\spa\bigl\{\ker(A^*-\gl):
\gl\in\bC\backslash\bR\bigr\}
\end{equation*}
and a self-adjoint operator $A_s$ in $\gotH\ominus\widehat\gotH$. Here
$\clospa \{\cdot\}$ denotes the closed linear span.
Obviously $A$ is simple if and only if $\widehat\gotH$ coincides
with $\gotH$. Notice that if $\Pi=\{\cH,\Gamma_0,\Gamma_1\}$ is a boundary triplet
for the adjoint $A^*$ of a non-simple symmetric operator $A=\widehat A\oplus A_s$, then
$\widehat\Pi=\{\cH,\widehat\Gamma_0,\widehat\Gamma_1\}$, where
\begin{equation*}
\widehat\Gamma_0:=\Gamma_0\upharpoonright\dom\bigl((\widehat A)^*\bigr)\quad
\text{and}\quad
\widehat\Gamma_1:=\Gamma_1\upharpoonright\dom\bigl((\widehat A)^*\bigr),
\end{equation*}
is a boundary triplet for the simple part
$(\widehat A)^*\in\cC(\widehat\gotH)$ such that the extension $A_\Theta=\Gamma^{(-1)}\Theta$, $\Theta\in
\widetilde\cC(\cH)$, in $\gotH$
is given by $\widehat A_\Theta\oplus A_s$, $\widehat A_\Theta:=\widehat\Gamma^{(-1)}\Theta\in\cC(\widehat\gotH)$,
and the Weyl functions and $\gamma$-fields of
$\Pi=\{\cH,\Gamma_0,\Gamma_1\}$ and $\widehat\Pi=\{\cH,\widehat\Gamma_0,\widehat\Gamma_1\}$
coincide.

We say that
a maximal dissipative operator is {\it completely non-self-adjoint}
if there is no nontrivial reducing subspace
in which it is self-adjoint. Notice that each
maximal dissipative operator decomposes orthogonally into a
self-adjoint part and a completely non-self-adjoint part, see e.g. \cite{FN}.

\subsection{Weyl functions, $\gamma$-fields and resolvents of extensions}\label{weylsec}

Let, as in Section \ref{btrips}, $A$ be a densely defined closed
symmetric operator in $\gotH$ with equal deficiency indices. If
$\lambda\in\dC$ is a point of regular type of $A$, i.e.
$(A-\lambda)^{-1}$ is bounded, we denote the {\it defect subspace}
of $A$ by $\kN_\gl=\ker(A^*-\gl)$. The following definition can
be found in \cite{DM87,DM91,DM95}.
\begin{defn}\label{Weylfunc}
Let $\Pi=\{\kH,\gG_0,\gG_1\}$ be a boundary triplet  for $A^*.$ The operator valued
functions
$\gamma(\cdot) :\ \rho(A_0)\rightarrow  [\kH,\gotH]$ and  $M(\cdot) :\
\rho(A_0)\rightarrow  [\kH]$ defined by
\begin{equation}\label{2.3A}
\gamma(\gl):=\bigl(\Gamma_0\!\upharpoonright\kN_\gl\bigr)^{-1} \qquad\text{and}\qquad
M(\gl):=\Gamma_1\gamma(\gl), \quad \gl\in\rho(A_0),
      \end{equation}
are called the {\em $\gamma$-field} and the {\em Weyl function}, respectively,
corresponding to the boundary triplet $\Pi.$
\end{defn}
It follows from the identity  $\dom(A^*)=\ker(\Gamma_0)\dot +\kN_\gl$,
$\lambda\in\rho(A_0)$, where as above $A_0=A^*\!\upharpoonright\ker(\gG_0)$,
that the $\gamma$-field  $\gamma(\cdot)$ and the Weyl function $M(\cdot)$ in \eqref{2.3A} are well defined.
Moreover both $\gamma(\cdot)$ and $M(\cdot)$ are
holomorphic on $\rho(A_0)$ and the relations
\begin{equation*}
\gamma(\lambda)=\bigl(I+(\gl-\mu)(A_0-\gl)^{-1}\bigr)\gamma(\mu),
\qquad \gl,\mu\in\rho(A_0),
\end{equation*}
and
\begin{equation}\label{mlambda}
M(\gl)-M(\mu)^*=(\gl-\overline\mu)\gamma(\mu)^*\gamma(\gl),
\qquad \gl,\mu\in\rho(A_0),
\end{equation}
are valid (see \cite{DM91}).
The identity \eqref{mlambda} yields that $M(\cdot)$ is a  {\it
Nevanlinna function}, that is, $M(\cdot)$ is a ($[\kH]$-valued)
holomorphic function on $\bC\backslash\dR$ and
\begin{equation}\label{mweyll}
M(\gl)=M(\overline\gl)^*\qquad\text{and}\qquad
\frac{\imag(M(\gl))}{\imag(\lambda)}\geq 0
\end{equation}
hold for all $\lambda\in\dC\backslash\dR$.
The union of $\dC\backslash\dR$ and the set of all points $\lambda\in\dR$ such that
$M$ can be analytically continued to $\lambda$ and the continuations from
$\dC_+$ and $\dC_-$ coincide is denoted by $\mathfrak h(M)$.
Besides \eqref{mweyll} it follows also from \eqref{mlambda}
that the Weyl function $M(\cdot)$ satisfies $0\in \rho(\imag(M(\gl)))$ for
all $\gl\in\bC\backslash\bR$; Nevanlinna functions with this additional property are sometimes
called uniformly strict, cf. \cite{DHMS06}.
Conversely, each $[\kH]$-valued Nevanlinna
function $\tau$ with the additional property $0\in\rho(\imag (\tau(\lambda)))$
for some (and hence for all) $\lambda\in\dC\backslash\dR$
can be realized as a Weyl function corresponding to some boundary triplet, we
refer to \cite{DM91,LT77,Mal92} for further details.

Let again $\Pi=\{\kH,\gG_0,\gG_1\}$ be a boundary triplet for $A^*$ with corresponding $\gamma$-field
$\gamma(\cdot)$ and Weyl function $M(\cdot)$. The spectrum and the resolvent set of the closed (not necessarily
self-adjoint) extensions of $A$ can be described with the help of the function $M(\cdot)$.
More precisely, if $A_\Theta\subseteq A^*$ is the extension
corresponding to $\Theta\in\widetilde\kC(\kH)$ via \eqref{bij}, then
a point $\gl\in\rho(A_0)$ belongs to
$\rho(A_\Theta)$ ($\sigma_i(A_\Theta)$, $i=p,c,r$) if and only if $0\in\rho(\Theta-M(\gl))$ (resp.
$0\in\sigma_i(\Theta-M(\gl))$, $i=p,c,r$). Moreover, for
$\gl\in\rho(A_0)\cap\rho(A_\Theta)$ the well-known resolvent formula
\begin{equation}\label{2.8}
(A_\Theta - \gl)^{-1} = (A_0 - \gl)^{-1} + \gga(\gl)\bigl(\Theta -
M(\gl)\bigr)^{-1}\gga(\overline{\gl})^*
\end{equation}
holds, cf. \cite{DM87,DM91,DM95}. Formula \eqref{2.8} is a generalization of the known Krein
formula for canonical resolvents. We emphasize that it is valid for
any closed extension $A_\Theta\subseteq A^*$ of $A$ with a nonempty resolvent set.

\subsection{Self-adjoint extensions and scattering}\label{scatsec}

Let $A$ be a densely defined closed symmetric operator in the separable Hilbert space
$\gotH$
and assume that the deficiency indices of $A$ coincide and are finite, i.e., $n_+(A)=n_-(A)<\infty$.
Let $\gP = \{\kH,\gG_0,\gG_1\}$, $A_0:=A^*\upharpoonright\ker(\Gamma_0)$,
be a boundary triplet for $A^*$ and let $A_\Theta$ be a self-adjoint extension of $A$
which corresponds to a self-adjoint $\Theta\in\widetilde\cC(\cH)$.
Since here $\dim\cH$ is finite by \eqref{2.8}
\begin{equation*}
(A_\Theta-\lambda)^{-1}-(A_0-\lambda)^{-1},\qquad\lambda\in\rho(A_\Theta)\cap\rho(A_0),
\end{equation*}
is a finite rank operator and therefore the pair $\{A_\gT,A_0\}$ performs a so-called
{\it complete scattering system}, that is, the {\it wave operators}
\begin{equation*}
W_\pm(A_\Theta,A_0) := \slim_{t\to\pm\infty}e^{itA_\Theta}e^{-itA_0}P^{ac}(A_0),
\end{equation*}
exist and their ranges coincide with the absolutely continuous
subspace $\gotH^{ac}(A_\Theta)$ of $A_\Theta$, cf. \cite{BW,Ka1,Wei1,Y}.
 $P^{ac}(A_0)$ denotes the orthogonal
projection onto the absolutely continuous subspace $\gotH^{ac}(A_0)$
of $A_0$.
The {\it scattering operator} $S(A_\Theta,A_0)$ of the {\it scattering system}
$\{A_\Theta,A_0\}$ is then defined by
\begin{equation*}
S(A_\Theta,A_0):= W_+(A_\Theta,A_0)^*W_-(A_\Theta,A_0).
\end{equation*}
If we regard the scattering operator as an operator in $\gotH^{ac}(A_0)$,
then $S(A_\Theta,A_0)$ is unitary, commutes with the absolutely continuous part
\begin{equation*}
A^{ac}_0:=A_0\upharpoonright \dom(A_0)\cap\gotH^{ac}(A_0)
\end{equation*}
of $A_0$ and it follows
that $S(A_\Theta,A_0)$ is unitarily equivalent to a multiplication operator
induced by a family $\{S_\Theta(\lambda)\}$ of unitary operators in
a spectral representation of $A_0^{ac}$, see e.g. \cite[Proposition 9.57]{BW}.
This family is called
the {\it scattering matrix} of the scattering system $\{A_\Theta,A_0\}$ and is one of the most
important quantities in the analysis of scattering processes.

We note that if the symmetric operator $A$ is not simple, then the Hilbert space $\gotH$
can be decomposed as $\gotH=\widehat\gotH\oplus(\widehat\gotH)^\bot$ (cf. the
end of Section~\ref{btrips}) such
that the scattering operator is given by the orthogonal sum $S(\widehat A_\Theta,\widehat A_0)\oplus I$,
where $A_\Theta=\widehat A_\Theta\oplus A_s$ and $A_0=\widehat A_0\oplus A_s$, and
hence it is sufficient to consider simple symmetric operators $A$ in the following.

Since the deficiency indices of $A$ are finite the Weyl function $M(\cdot)$
corresponding to the boundary triplet $\gP = \{\kH,\gG_0,\gG_1\}$ is
a matrix-valued Nevanlinna function. By Fatous theorem (see \cite{Don,Gar}) then the limit
\begin{equation}\label{mlim}
M(\gl + i0) := \lim_{\epsilon\to+0}M(\gl + i\epsilon)
\end{equation}
from the upper half-plane exists for a.e. $\gl \in \bR$. We denote the set of real points
where the limit in \eqref{mlim} exits by $\gS^M$ and we agree to use a similar
notation for arbitrary scalar and matrix-valued Nevanlinna functions. Furthermore we will make use of the notation
\begin{equation}\label{hm}
\kH_{M(\gl)} := \ran\bigl(\imag(M(\gl))\bigr),\qquad\lambda\in\Sigma^M,
\end{equation}
and we will in general regard $\cH_{M(\lambda)}$ as a subspace of $\cH$.
The orthogonal projection and restriction onto $\cH_{M(\lambda)}$ will be denoted by $P_{M(\lambda)}$
and $\upharpoonright_{\cH_{M(\lambda)}}$, respectively.
Notice that for $\lambda\in\rho(A_0)\cap\dR$ the Hilbert space $\kH_{M(\gl)}$
is trivial by \eqref{mlambda}. Again we agree to use a notation analogous to \eqref{hm}
for arbitrary Nevanlinna functions. The family $\{P_{M(\lambda)}\}_{\lambda\in\Sigma^M}$ of orthogonal
projections in $\cH$ onto $\cH_{M(\lambda)}$, $\lambda\in\Sigma^M$, is measurable and
defines an orthogonal projection in the Hilbert space $L^2(\dR,d\lambda,\cH)$; sometimes we write
$L^2(\dR,\cH)$ instead of $L^2(\dR,d\lambda,\cH)$. The range of
this projection is denoted by $L^2(\dR,d\lambda,\cH_{M(\lambda)})$.

Besides the Weyl function $M(\cdot)$ we will also make use of the function
\begin{equation}\label{ntheta}
\lambda\mapsto N_\Theta(\lambda):=\bigl(\Theta-M(\lambda)\bigr)^{-1},\qquad\lambda\in\dC\backslash\dR,
\end{equation}
where $\Theta\in\widetilde\cC(\cH)$ is the self-adjoint relation corresponding to the extension $A_\Theta$
via \eqref{bij}. Since $\lambda\in\rho(A_0)\cap\rho(A_\Theta)$ if and only if $0\in\rho(\Theta-M(\lambda))$
the function $N_\Theta(\cdot)$ is well defined. It is not difficult to see
that $N_\Theta(\cdot)$ is an $[\cH]$-valued Nevanlinna function
and hence $N_{\Theta}(\lambda+i0)=
\lim_{\epsilon\rightarrow 0}N_{\Theta}(\lambda+i\epsilon)$ exists for almost every
$\lambda\in\dR$, we denote this set by $\Sigma^{N_{\Theta}}$.
We claim that
\begin{equation}\label{gvn}
N_{\Theta}(\lambda+i0)=\bigl(\Theta-M(\lambda+i0)\bigr)^{-1},\qquad\lambda\in\Sigma^M\cap\Sigma^{N_\Theta},
\end{equation}
holds. In fact, if $\Theta$ is a self-adjoint matrix then \eqref{gvn} follows immediately
from
$N_\Theta(\lambda)(\Theta-M(\lambda))=(\Theta-M(\lambda))N_\Theta(\lambda)=I_\cH$, $\lambda\in\dC_+$.
If $\Theta\in\widetilde\cC(\cH)$ has
a nontrivial multivalued part we
decompose $\Theta$ as
$\Theta=\Theta_{\rm op}\oplus\Theta_{\infty}$, where $\Theta_{\rm op}$ is
a self-adjoint matrix in $\cH_{\rm op}=\dom\Theta_{\rm op}$ and $\Theta_{\infty}$ is a pure relation in
$\cH_\infty=\cH\ominus\cH_{\rm op}$, cf. Section~\ref{btrips}, and
denote the orthogonal projection and restriction in $\cH$ onto $\cH_{\rm op}$ by $P_{\rm op}$
and $\upharpoonright_{\cH_{\rm op}}$, respectively. Then we have
\begin{equation*}
\lambda\mapsto N_{\Theta}(\lambda)=\bigl(\Theta_{\rm op}-P_{\rm op}M(\lambda)
\!\upharpoonright_{\cH_{\rm op}}\bigr)^{-1}P_{\rm op},\qquad\lambda\in\dC\backslash\dR,
\end{equation*}
(see e.g. \cite[page~137]{LT77}) and from
 $N_{\Theta}(\lambda+i0)=(\Theta_{\rm op}-P_{\rm op}M(\lambda+i0)\!
\upharpoonright_{\cH_{\rm op}})^{-1}P_{\rm op}$ for all $\lambda\in\Sigma^M\cap\Sigma^{N_\Theta}$
we conclude \eqref{gvn}. Notice that the set $\dR\backslash (\Sigma^M\cap\Sigma^{N_\Theta})$ has Lebesgue measure zero.

The following representation theorem of
the scattering matrix $\{S_\gT(\gl)\}_{\gl \in \bR}$ of the scattering
system $\{A_\gT,A_0\}$ is essential in the following, cf.
\cite[Theorem~3.8]{BMN}. Since the scattering matrix is only determined up
to a set of Lebesgue measure zero we choose the representative
of the equivalence class defined on $\Sigma^M\cap\Sigma^{N_\Theta}$.

\begin{thm}\label{scattering}
Let $A$ be a densely defined closed simple symmetric operator with
finite deficiency indices in the separable Hilbert space $\gotH$,
let $\Pi= \{\kH,\Gamma_0,\Gamma_1\}$ be a boundary triplet for $A^*$
with corresponding Weyl function $M(\cdot)$ and define $\cH_{M(\lambda)}$,
$\lambda\in\Sigma^M$, as in \eqref{hm}. Furthermore, let
$A_0=A^*\!\upharpoonright\ker(\Gamma_0)$  and
let $A_\Theta=A^*\upharpoonright\Gamma^{(-1)}\Theta$,
$\Theta\in\widetilde\kC(\kH)$, be a self-adjoint extension of $A$.
Then the following holds.
\begin{itemize}
\item [{\rm (i)}]
$A^{ac}_0$ is unitarily equivalent to the multiplication operator
with the free variable in $L^2(\bR,d\gl,\kH_{M(\gl)})$.
\item [{\rm (ii)}] In $L^2(\bR,d\gl,\kH_{M(\gl)})$ the scattering matrix $\{S_\Theta(\gl)\}$
of the complete scattering system $\{A_\Theta,A_0\}$ is given by
{\small
\begin{displaymath}
S_\Theta(\gl) = I_{\kH_{M(\gl)}} +
2iP_{M(\lambda)}\sqrt{\imag(M(\gl))}\bigl(\Theta-M(\gl)\bigr)^{-1}
\sqrt{\imag(M(\gl))}\upharpoonright_{\cH_{M(\lambda)}}
\end{displaymath}
}
for all $\gl \in\Sigma^M\cap\Sigma^{N_\Theta}$, where $M(\gl):=M(\gl + i0)$.
\end{itemize}
\end{thm}
In order to show the usefulness of Theorem~\ref{scattering} and to make the reader more familiar with the notion of boundary triplets and
associated Weyl functions we calculate the scattering matrix of the scattering system
$\{-\tfrac{d^2}{dx^2}+\delta,-\tfrac{d^2}{dx^2}\}$ in the following simple example.

\begin{exam}\label{delta}
{\rm
Let us consider the densely defined closed simple symmetric
operator
\begin{equation*}
(Af)(x):=-f^{\prime\prime}(x),\quad\dom(A)=\bigl\{f\in W^2_2(\dR):f(0)=0\bigr\},
\end{equation*}
in $L^2(\dR)$, see e.g. \cite{AGHH1}. Clearly $A$ has deficiency indices $n_+(A)=n_-(A)=1$ and it is well-known
that the adjoint operator $A^*$ is given by $(A^*f)(x)=-f^{\prime\prime}(x)$,
\begin{equation*}
\dom(A^*)=\bigr\{f\in W^2_2(\dR\backslash\{0\}):f(0+)=f(0-),\,f^{\prime\prime}\in L^2(\dR)\bigl\}.
\end{equation*}
It is not difficult to verify that $\Pi=\{\dC,\Gamma_0,\Gamma_1\}$, where
\begin{equation*}
\Gamma_0f:=f^\prime(0+)-f^\prime(0-)\quad\text{and}\quad\Gamma_1 f:=-f(0+),\quad f\in\dom(A^*),
\end{equation*}
is a boundary triplet for $A^*$ and $A_0=A^*\upharpoonright\ker(\Gamma_0)$ coincides with the
usual self-adjoint second order differential operator defined on $W^2_2(\dR)$. Moreover the defect space
$\ker(A^*-\lambda)$,
$\lambda\not\in [0,\infty)$, is spanned by the function
\begin{equation*}
x\mapsto e^{i\sqrt{\lambda}x}\chi_{\dR_+}(x)+
e^{-i\sqrt{\lambda}x}\chi_{\dR_-}(x),\quad \lambda\not\in [0,\infty),
\end{equation*}
where the square root is defined on $\bC$ with a cut along $[0,\infty)$ and fixed by
$\imag(\sqrt{\gl})>0$ for $\lambda\not\in [0,\infty)$ and by $\sqrt{\lambda}\geq 0$ for
$\lambda\in[0,\infty)$. Therefore we find that the Weyl function $M(\cdot)$ corresponding to
$\Pi=\{\dC,\Gamma_0,\Gamma_1\}$ is given by
\begin{equation*}
M(\lambda)=\frac{\Gamma_1 f_\lambda}{\Gamma_0 f_\lambda}=\frac{i}{2\sqrt{\lambda}},\qquad f_\lambda\in
\ker(A^*-\lambda),\quad \lambda\not\in [0,\infty).
\end{equation*}
Let $\alpha\in\dR\backslash\{0\}$ and consider the self-adjoint extension
$A_{-\alpha^{-1}}$ corresponding to the parameter $-\alpha^{-1}$, $A_{-\alpha^{-1}}=A^*\upharpoonright\ker(\Gamma_1+\alpha^{-1}\Gamma_0)$, i.e.
\begin{equation*}
\begin{split}
(A_{-\alpha^{-1}}f)(x)&=-f^{\prime\prime}(x)\\
\dom(A_{-\alpha^{-1}})&=
\bigl\{f\in\dom(A^*):\alpha f(0\pm)=f^\prime(0+)-f^\prime(0-)\bigr\}.
\end{split}
\end{equation*}
This self-adjoint operator is often denoted by
$-\tfrac{d^2}{dx^2}+\alpha\delta$, see \cite{AGHH1}. It follows immediately
from Theorem~\ref{scattering} that the scattering matrix $\{S(\lambda)\}$ of the scattering system
$\{A_{-\alpha^{-1}},A_0\}$ is given by
\begin{equation*}
S(\lambda)=\frac{2\sqrt{\gl} - i\ga}{2\sqrt{\gl} + i\ga},\qquad\lambda>0.
\end{equation*}
We note that scattering systems of the form
$\{-\tfrac{d^2}{dx^2}+\alpha\delta^\prime,-\tfrac{d^2}{dx^2}\}$, $\alpha\in\dR$, can be investigated in
a similar way as above. Other examples can be found in \cite{BMN}.
}
\end{exam}

\section{Dissipative and Lax-Phillips scattering systems}\la{dilations}

In this section we regard scattering systems $\{A_D,A_0\}$ consisting of a
maximal dissipative and a self-adjoint extension of a symmetric operator $A$
with finite deficiency indices. In the theory of open quantum system the maximal
dissipative operator $A_D$ is often called a pseudo-Hamiltonian. We shall explicitely
construct a dilation (or so-called quasi-Hamiltonian) $\widetilde K$ of $A_D$ and calculate
the scattering matrix of the scattering system $\{\widetilde K,A_0\oplus G_0\}$, where $G_0$
is a self-adjoint first order differential operator. The diagonal entries of the scattering
matrix then turn out to be the scattering matrix of the dissipative scattering system
$\{A_D,A_0\}$ and of a so-called Lax-Phillips
scattering system, respectively.

We emphasize that this efficient and somehow straightforward method for the analysis of
scattering processes for open quantum systems has the essential disadvantage that
the quasi-Hamiltonians $\widetilde K$ and $A_0\oplus G_0$ are necessarily not semibounded from below.

\subsection{Self-adjoint dilations of maximal dissipative operators}\label{sec3.1}

Let in the following  $A$ be a densely defined closed
simple symmetric operator in the separable Hilbert space
$\gotH$ with equal finite deficiency indices $n_\pm(A)=n<\infty$,
let $\gP = \{\kH,\gG_0,\gG_1\}$, $A_0=A^*\upharpoonright\ker(\Gamma_0)$, be a boundary triplet for $A^*$
and let $D \in [\kH]$ be a dissipative $n\times n$-matrix.
Then the closed extension
\begin{equation*}
A_D=A^*\upharpoonright\ker(\Gamma_1-D\Gamma_0)
\end{equation*}
of $A$ corresponding to $\Theta = D$ via \eqref{bij}-\eqref{bijop} is maximal
dissipative and $\bC_+$ belongs to $\rho(A_D)$. Notice that here we restrict ourselves to
maximal dissipative extensions $A_D$ corresponding to dissipative matrices $D$ instead of maximal dissipative
relations in the finite dimensional space $\cH$.
This is no essential restriction, see Remark~\ref{remdis} at the end of this subsection.
For $\gl \in \rho(A_D)\cap\rho(A_0)$ the resolvent of
the extension $A_D$ is given by
\begin{equation}\la{3.1}
(A_D -\gl)^{-1} = (A_0 - \gl)^{-1} + \gga(\gl)\bigl(D -
M(\gl)\bigr)^{-1}\gga(\overline{\gl})^*,
\end{equation}
cf. \eqref{2.8}. Write the dissipative matrix $D\in[\cH]$ as
\begin{equation*}
D=\real (D)+i\imag (D),
\end{equation*}
decompose $\cH$ as the direct orthogonal sum of the finite dimensional
subspaces $\ker(\imag(D))$ and $\cH_D:=\ran(\imag(D))$,
\begin{equation}\label{decoh}
\cH=\ker(\imag(D))\oplus\cH_D,
\end{equation}
and denote by $P_D$ and $\upharpoonright_{\cH_D}$ the orthogonal projection and restriction
in $\cH$ onto $\cH_D$. Since $\imag(D)\leq 0$ the self-adjoint matrix
$-P_D\imag(D)\!\upharpoonright_{\cH_D}\in[\cH_D]$ is strictly positive and the next lemma shows how
$-iP_D\imag(D)\!\upharpoonright_{\cH_D}$ (and $iP_D\imag(D)\!\upharpoonright_{\cH_D}$) can
be realized as a Weyl function of a differential operator.

\begin{lem}\label{disslem}
Let
$G$ be the symmetric first order differential operator in the Hilbert
space $L^2(\dR,\cH_D)$ defined by
\begin{equation*}
(Gg)(x)=-ig^\prime(x),\qquad\dom (G)=\bigl\{g\in W^1_2(\dR,\cH_D)\,:\,g(0)=0\bigr\}.
\end{equation*}
Then $G$ is simple, $n_\pm(G)=\dim\cH_D$ and the adjoint operator
$G^*g=-ig^\prime$ is defined on $\dom (G^*)=W^1_2(\dR_-,\cH_D)\oplus W^1_2(\dR_+,\cH_D)$.
Moreover, the triplet $\Pi_G=\{\cH_D,\gY_0,\gY_1\}$,
where
\begin{equation*}
\begin{split}
\gY_0 g&:=\frac{1}{\sqrt{2}}\bigl(-P_D\imag(D)\!\upharpoonright_{\cH_D}\bigr)^{-\frac{1}{2}}\bigl(g(0+)-g(0-)\bigr),\\
\gY_1 g&:=\frac{i}{\sqrt{2}}\bigl(-P_D\imag(D)\!\upharpoonright_{\cH_D}\bigr)^{\frac{1}{2}}
\bigl(g(0+)+g(0-)\bigr),
\end{split}
\end{equation*}
$g\in\dom(G^*)$, is a boundary triplet for $G^*$ and $G_0:=G^*\upharpoonright\ker(\gY_0)$ is the usual
self-adjoint first order differential operator in $L^2(\dR,\cH_D)$ with domain $\dom(G_0)=W^{1}_2(\dR,\cH_D)$
and $\sigma(G_0)=\dR$.
The Weyl function $\tau(\cdot)$ corresponding to the boundary triplet $\Pi_G=\{\cH_D,\gY_0,\gY_1\}$ is
given by
\begin{equation}\label{taud}
\tau(\lambda)=\begin{cases} -iP_D\imag(D)\!\upharpoonright_{\cH_D}, & \lambda\in\dC_+,\\
i P_D\imag(D)\!\upharpoonright_{\cH_D}, & \lambda\in\dC_-.\end{cases}
\end{equation}
\end{lem}

\begin{proof}
Besides the assertion that $\Pi_G=\{\cH_D,\gY_0,\gY_1\}$ is a boundary triplet for $G^*$ with
Weyl function $\tau(\cdot)$ given by \eqref{taud} the statements of the lemma are well-known.
We note only that the simplicity of $G$ follows from \cite[VIII.104]{AG} and
the fact that $G$ can be written as a finite direct
orthogonal sum of first order differential operators on $\dR_-$ and $\dR_+$.

A straightforward calculation shows that the identity
\begin{equation*}
\begin{split}
(G^*g,k)-(g,G^*k)&=i(g(0+),k(0+))-i(g(0-),k(0-))\\
&=(\gY_1 g,\gY_0 k)-(\gY_0 g,\gY_1 k)
\end{split}
\end{equation*}
holds for all $g,k\in\dom(G^*)$. Moreover the mapping $(\Gamma_0,\Gamma_1)^\top$ is surjective.
Indeed, for an element $(h,h^\prime)^\top\in\cH_D\times\cH_D$ we choose $g\in\dom G^*$
such that
\begin{equation*}
g(0+)=\frac{1}{\sqrt{2}}\Bigl\{\bigl(-P_D\imag(D)\!\upharpoonright_{\cH_D}\bigr)^{\frac{1}{2}} h
-i \bigl(-P_D\imag(D)\!\upharpoonright_{\cH_D}\bigr)^{-\frac{1}{2}} h^\prime \Bigr\}
\end{equation*}
and
\begin{equation*}
g(0-)=\frac{1}{\sqrt{2}}\Bigl\{-\bigl(-P_D\imag(D)\!\upharpoonright_{\cH_D}\bigr)^{\frac{1}{2}} h
-i \bigl(-P_D\imag(D)\!\upharpoonright_{\cH_D}\bigr)^{-\frac{1}{2}} h^\prime \Bigr\}
\end{equation*}
holds. Then a simple calculation shows $\gY_0 g=h$, $\gY_1g=h^\prime$ and
therefore $\Pi_G=\{\cH_D,\gY_0,\gY_1\}$
is a boundary triplet for $G^*$.
It is not difficult to check that the defect subspace $\cN_\lambda=\ker(G^*-\lambda)$ is
\begin{equation*}
\cN_\lambda=\begin{cases}\sp\bigl\{x\mapsto e^{i\lambda x}\chi_{\dR_+}(x)\xi\,:\,\xi\in\cH_D\bigr\},
&\lambda\in\dC_+,\\
\sp\bigl\{x\mapsto e^{i\lambda x}\chi_{\dR_-}(x)\xi\,:\,\xi\in\cH_D\bigr\},
&\lambda\in\dC_-,
\end{cases}
\end{equation*}
and hence we conclude that the Weyl function of $\Pi_G=\{\cH_D,\gY_0,\gY_1\}$ is given by \eqref{taud}.
\end{proof}

Let $A_D$ be the maximal dissipative extension of $A$ in $\gotH$ from above and let $G$
be the first order differential operator from Lemma~\ref{disslem}.
Clearly
$K := A \oplus G$ is a densely defined closed simple symmetric operator in the separable
Hilbert space
\begin{equation*}
\gotK:=\gotH\oplus L^2(\dR,\cH_D)
\end{equation*}
with equal finite deficiency indices $n_\pm(K)=n_\pm(A)+n_\pm(G)<\infty$ and the adjoint
is $K^*=A^*\oplus G^*$. The elements in $\dom(K^*)=\dom(A^*)\oplus\dom(G^*)$ will be
written in the form $f\oplus g$, $f\in\dom(A^*)$, $g\in\dom (G^*)$. In the next theorem
we construct a self-adjoint extension $\widetilde K$ of $K$ in $\gotK$ which is a minimal
self-adjoint dilation of the dissipative operator $A_D$ in $\gotH$. 
The construction is based on the idea of the coupling method from \cite{DHMS00}. It is
worth to mention that in the case of a (scalar) Sturm-Liouville operator with real potential and
dissipative boundary condition our construction coincides with the one proposed by 
B.S.~Pavlov \cite{Pa1}, cf. Example~\ref{pavlov} below.

\begin{thm}\label{III.1}
Let $A$, $\Pi=\{\cH,\Gamma_0,\Gamma_1\}$ and $A_D$ be as in the beginning of this section,
let $G$ and $\Pi_G=\{\cH_D,\gY_0,\gY_1\}$
be as in Lemma~{\rm \ref{disslem}} and $K=A\oplus G$. Then
\begin{equation}\label{widetildek}
\widetilde K=K^*\upharpoonright\left\{f\oplus g\in\dom(K^*):
\begin{matrix}
P_D\Gamma_0f-\gY_0 g=0,\\(1-P_D)(\Gamma_1-\real(D)\Gamma_0)f=0,\\
P_D(\Gamma_1-\real(D)\Gamma_0)f+\gY_1 g=0
\end{matrix}\right\}
\end{equation}
is a minimal self-adjoint dilation of the maximal dissipative
operator $A_D$, that is, for all $\lambda\in\dC_+$
\begin{equation*}
P_\gotH\bigl(\widetilde K-\lambda\bigr)^{-1}\upharpoonright_{\gotH}=(A_D-\lambda)^{-1}
\end{equation*}
holds and the minimality condition
$\gotK=\clospa\{(\widetilde K-\lambda)^{-1}\gotH:\lambda\in\dC\backslash\dR\}$
is satisfied. Moreover $\sigma(\widetilde K)=\dR$.
\end{thm}

\begin{proof}
Let $\gamma(\cdot),\nu(\cdot)$ and $M(\cdot),\tau(\cdot)$ be the $\gamma$-fields and Weyl functions
of the boundary triplets $\Pi=\{\cH,\Gamma_0,\Gamma_1\}$ and $\Pi_G=\{\cH_D,\gY_0,\gY_1\}$,
respectively. Then it is straightforward to check that
$\widetilde{\gP}=
\{\widetilde{\kH},\widetilde{\gG}_0,\widetilde{\gG}_1\}$, where
\begin{equation}\la{3.12}
\widetilde{\kH} := \kH \oplus \kH_D, \quad \widetilde{\gG}_0 :=
\begin{pmatrix}\gG_0\\ \gY_0\end{pmatrix}\quad\text{and} \quad \widetilde{\gG}_1 :=
\begin{pmatrix}\gG_1-\real(D)\Gamma_0\\ \gY_1\end{pmatrix},
\end{equation}
is a boundary triplet for $K^* = A^* \oplus G^*$ and the corresponding Weyl function
$\widetilde M(\cdot)$ and $\gamma$-field $\widetilde\gamma(\cdot)$ are given by
\begin{equation}\la{3.14a}
\widetilde{M}(\gl) =
\begin{pmatrix}
M(\gl)-\real(D) & 0\\
0 & \gt(\gl)
\end{pmatrix}, \qquad \gl \in \dC\backslash\dR,
\end{equation}
and
\begin{equation}\label{3.14aa}
\widetilde{\gga}(\gl) =
\begin{pmatrix}
\gga(\gl) & 0\\
0 & \nu(\gl)
\end{pmatrix} , \qquad \gl \in \dC\backslash\dR,
\end{equation}
respectively. Notice also that $K_0:=K^*\upharpoonright\ker(\widetilde\Gamma_0)=A_0\oplus G_0$
holds.

With respect to the decomposition $\widetilde{\kH} = \ker(\imag(D)) \oplus \kH_D \oplus
\kH_D$ of $\widetilde\cH$ (cf. \eqref{decoh}) we define the linear relation $\widetilde{\gT}$ by
\begin{equation}\label{wtdef}
\widetilde\Theta:=\left\{\begin{pmatrix}(u,v,v)^\top\\ (0,-w,w)^\top\end{pmatrix}:
u\in\ker(\imag(D),\; v,w\in\cH_D\right\}
\in\widetilde\cC(\widetilde\cH).
\end{equation}
We leave it to the reader to check that $\widetilde\Theta$ is self-adjoint.
Hence by Proposition~\ref{propo} the operator $K_{\widetilde\Theta}=
K^*\upharpoonright\widetilde\Gamma^{(-1)}\widetilde\Theta$
is a self-adjoint extension of the symmetric operator $K=A\oplus G$ in $\gotK=\gotH\oplus L^2(\dR,\cH_D)$
and one verifies without difficulty that this extension coincides with $\widetilde K$ from
\eqref{widetildek}, $\widetilde K=K_{\widetilde\Theta}$.

In order to calculate $(\widetilde K-\lambda)^{-1}$, $\lambda\in\dC\backslash\dR$,
we use the block matrix decomposition
\begin{equation}\label{mdreal}
M(\gl)-\real(D) =
\begin{pmatrix}
M^D_{11}(\gl) & M^D_{12}(\gl)\\
M^D_{21}(\gl) & M^D_{22}(\gl)
\end{pmatrix}\in
\bigl[\ker(\imag(D))\oplus\cH_D\bigr]
\end{equation}
of $M(\lambda)-\real(D)\in[\cH]$.
Then the definition of $\widetilde\Theta$ in \eqref{wtdef} and \eqref{3.14a} imply
\begin{equation*}
\bigl(\widetilde\Theta-\widetilde M(\lambda)\bigr)^{-1}=\left\{\begin{pmatrix}
\begin{pmatrix} -M^D_{11}(\gl)u -
M^D_{12}(\gl)v\\ -w - M^D_{21}(\gl)u - M^D_{22}(\gl)v \\w - \gt(\gl) v\end{pmatrix}
\\(u,v,v)^\top
\end{pmatrix}: \begin{matrix}
u\in\ker(\imag(D)\\ v,w\in\cH_D\end{matrix}\right\}
\end{equation*}
and since every $\lambda\in\dC\backslash\dR$ belongs to $\rho(\widetilde K)\cap\rho(K_0)$,
$K_0=A_0\oplus G_0$, it follows that
$(\widetilde\Theta-\widetilde M(\lambda))^{-1}$, $\lambda\in\dC\backslash\dR$, is the graph of
a bounded everywhere defined operator. In order to calculate $(\widetilde\Theta-\widetilde M(\lambda))^{-1}$
in a more explicit form we set
\begin{equation}\label{xyz}
\ba{rcl}
x & := & -M^D_{11}(\gl)u - M^D_{12}(\gl)v, \\
y & := & -w - M^D_{21}(\gl)u - M^D_{22}(\gl)v, \\
z & := & w - \gt(\gl) v.
\ea
\end{equation}
This yields
\begin{equation*}
\begin{pmatrix}
x \\
y + z
\end{pmatrix} =
-\begin{pmatrix}
M^D_{11}(\gl) & M^D_{12}(\gl)\\
M^D_{21}(\gl) & M^D_{22}(\gl) + \gt(\gl)
\end{pmatrix}
\begin{pmatrix}
u\\
v
\end{pmatrix}
\end{equation*}
and by \eqref{taud} and \eqref{mdreal} we have
\begin{equation}\label{mdweyl}
-\begin{pmatrix}
M^D_{11}(\gl) & M^D_{12}(\gl)\\
M^D_{21}(\gl) & M^D_{22}(\gl) + \gt(\gl)
\end{pmatrix}=\begin{cases} D-M(\lambda), & \lambda\in\dC_+,\\ D^*-M(\lambda), & \lambda\in\dC_-
\end{cases}.
\end{equation}
Hence for $\lambda\in\dC_+$ we find
\begin{equation*}
\begin{pmatrix}
u\\
v
\end{pmatrix}=
\bigl(D-M(\lambda)\bigr)^{-1}
\begin{pmatrix}
x \\
y + z
\end{pmatrix},
\end{equation*}
which implies
\begin{equation}\label{uv}
\begin{pmatrix}
u\\
v
\end{pmatrix}=
\bigl(D-M(\lambda)\bigr)^{-1}
\begin{pmatrix}
x \\
y
\end{pmatrix}+ \bigl(D-M(\lambda)\bigr)^{-1}\upharpoonright_{\cH_D} z
\end{equation}
and
\begin{equation}\label{uv1}
v=
P_D\bigl(D-M(\lambda)\bigr)^{-1}
\begin{pmatrix}
x \\
y
\end{pmatrix}+ P_D\bigl(D-M(\lambda)\bigr)^{-1}\upharpoonright_{\cH_D} z.
\end{equation}
Therefore by inserting \eqref{xyz}, \eqref{uv} and \eqref{uv1} into the above expression
for $(\widetilde\Theta-\widetilde M(\lambda))^{-1}$ we obtain
\begin{equation}\la{3.27}
\bigl(\widetilde{\gT} - \widetilde{M}(\gl)\bigr)^{-1} =
\begin{pmatrix}
(D - M(\gl))^{-1} & (D - M(\gl))^{-1}\upharpoonright_{\cH_D}\\
P_D(D - M(\gl))^{-1} & P_D(D - M(\gl))^{-1}\upharpoonright_{\cH_D}
\end{pmatrix}
\end{equation}
for all $\gl \in \bC_+$ and by \eqref{2.8} the resolvent of the self-adjoint extension $\widetilde K$
admits the representation
\begin{equation}\label{resok}
\bigl(\widetilde K - \gl\bigr)^{-1} = (K_0 - \gl)^{-1} +
\widetilde{\gga}(\gl)\bigl(\widetilde{\gT} - \widetilde{M}(\gl)\bigr)^{-1}
\widetilde{\gga}(\overline{\gl})^*,
\end{equation}
$\gl \in \dC\backslash\dR$. It follows from $K_0=A_0\oplus G_0$, \eqref{3.14aa} and \eqref{3.27}
that for $\lambda\in\dC_+$ the compressed resolvent of $\widetilde K$ onto $\gotH$ is given by
\begin{equation*}
P_\gotH\bigl(\widetilde K - \gl\bigr)^{-1}\upharpoonright\gotH
  = (A_0 - \gl)^{-1} + \gga(\gl)\bigl(D -
  M(\gl)\bigr)^{-1}\gga(\overline{\gl})^*,
\end{equation*}
where $P_\gotH$ denotes the orthogonal projection in $\gotK$ onto $\gotH$.
Taking into account \eqref{3.1} we get
\begin{equation*}
P_\gotH\bigl(\widetilde K -
\gl\bigr)^{-1}\upharpoonright\gotH = (A_D - \gl)^{-1},
\quad \gl \in \bC_+,
\end{equation*}
and hence $\widetilde K$ is a self-adjoint dilation of $A_D$. Since $\sigma(G_0)=\dR$ it follows
from well-known perturbation results and \eqref{resok} that $\sigma(\widetilde K)=\dR$ holds.

It remains to show that $\widetilde K$ satisfies the minimality condition
\begin{equation}\label{miniq}
\gotK=\gotH\oplus L^2(\dR,\cH_D)=\clospa\bigl\{(\widetilde K-\lambda)^{-1}\gotH:\lambda\in\dC\backslash\dR\bigr\}.
\end{equation}
First of all $\slim_{t\to+\infty}(-it)(\widetilde K - it)^{-1} =
I_{\gotK}$ implies that $\gotH$ is a subset of the right hand side of \eqref{miniq}.
The orthogonal projection in $\gotK$ onto $L^2(\dR,\cH_D)$ is denoted by $P_{L^2}$.
Then we conclude from \eqref{3.14aa}, \eqref{3.27} and \eqref{resok} that for $\lambda\in\dC_+$
\begin{equation}\label{compl2}
P_{L^2}\bigl(\widetilde K-\lambda\bigr)^{-1}\!\upharpoonright_{\gotH}=
\nu(\gl)P_D\bigl(D - M(\gl)\bigr)^{-1}\gga(\overline{\gl})^*
\end{equation}
holds and this gives
\begin{equation*}
\ran\bigl(P_{L^2}\bigl(\widetilde K-\lambda\bigr)^{-1}\!\upharpoonright_{\gotH}\bigr)
=\ker(G^*-\lambda),\quad  \lambda\in\dC_+.
\end{equation*}
From \eqref{mdweyl} it follows that similar to the matrix
representation \eqref{3.27}
the left lower corner of $(\widetilde\Theta-\widetilde M(\lambda))^{-1}$
is given by $P_D(D^*-M(\lambda))^{-1}$ for $\lambda\in\dC_-$. Hence,
the analogon of \eqref{compl2} for $\lambda\in\dC_-$  implies that
\begin{equation*}
\ran\bigl(P_{L^2}\bigl(\widetilde K-\lambda\bigr)^{-1}\!\upharpoonright_{\gotH}\bigr)
=\ker(G^*-\lambda)
\end{equation*}
is true for $\lambda\in\dC_-$. Since by Lemma~\ref{disslem} the symmetric operator $G$ is simple
it follows that
\begin{equation*}
L^2(\dR,\cH_D)=\clospa\bigl\{\ker(G^*-\lambda):\lambda\in\dC\backslash\dR\bigr\}
\end{equation*}
holds, cf. Section~\ref{btrips},
and therefore the minimality condition \eqref{miniq} holds.
\end{proof}

\begin{rem}\label{remdis}
{\rm
We note that also in the case where the parameter $D$ is not a dissipative matrix
but a maximal dissipative relation in $\cH$ a minimal self-adjoint dilation of $A_D$ can
be constructed in a similar way as in Theorem~\ref{III.1}.

Indeed, let $A$ and $\Pi=\{\cH,\Gamma_0,\Gamma_1\}$ be as in the beginning of this
section and let $\widetilde D\in\widetilde\cC(\cH)$ be a maximal dissipative relation in
$\cH$. Then $\widetilde D$ can be written as the direct orthogonal sum of a dissipative
matrix $\widetilde D_{\rm op}$ in $\cH_{\rm op}:=\cH\ominus\mul\widetilde D$ and an
undetermined part  or "pure relation"
$\widetilde D_\infty := \{\bigl(\begin{smallmatrix} 0\\ y\end{smallmatrix}\bigr):\ y\in \mul\widetilde D$\}. 
It follows that
\begin{equation*}
B:=A^*\upharpoonright\Gamma^{(-1)}\bigl\{\bigl(\begin{smallmatrix} 0\\ y\end{smallmatrix}\bigr):
y\in\mul\widetilde D\bigr\}=A^*\upharpoonright\Gamma^{(-1)}\widetilde D_\infty
\end{equation*}
is a closed symmetric extension of $A$ and 
$\{\cH_{\rm op},\Gamma_0\!\upharpoonright_{\dom(B^*)},P_{\rm op}\Gamma_1\!\upharpoonright_{\dom(B^*)}\}$ is
a boundary triplet for
\begin{equation*}
B^*=A^*\upharpoonright\bigl\{f\in\dom(A^*):(1-P_{\rm op})\Gamma_0f=0\bigr\}
\end{equation*}
with $A^*\!\upharpoonright\ker(\Gamma_0)=B^*\!\upharpoonright\ker(\Gamma_0\!\upharpoonright_{\dom(B^*)})$.
In terms of this boundary triplet the maximal dissipative extension $A_{\widetilde D}=\Gamma^{(-1)}\widetilde D$
coincides with the extension
\begin{equation*}
B_{\widetilde D_{\rm op}}=
B^*\upharpoonright\ker\bigl(P_{\rm op}\Gamma_1\!\upharpoonright_{\dom(B^*)}-
\widetilde D_{\rm op}\Gamma_0\!\upharpoonright_{\dom(B^*)}\bigr)
\end{equation*}
corresponding to the operator part $\widetilde D_{\rm op}\in[\cH_{\rm op}]$ of $\widetilde D$.
}
\end{rem}
\begin{rem}\label{specrem}
{\rm 
In the special case $\ker(\imag D)=\{0\}$ the relations \eqref{widetildek} take the form
\begin{equation*}
\Gamma_0f- \gY_0 g=0\quad\text{and}\quad  (\Gamma_1-\real(D)\Gamma_0)f+\gY_1 g=0,
\end{equation*}
so that $\widetilde K$ is a coupling of the self-adjoint operators $A_0$  and $G_0$ 
corresponding to the coupling of the boundary
triplets $\Pi_A=\{{\mathcal H},\Gamma_0,\Gamma_1-\real(D)\Gamma_0\}$ and  $
\Pi_G=\{{\mathcal H},\gY_0,\gY_1\}$ in the sense of \cite{DHMS00}.
In the case $\ker (\imag D)\not =\{0\}$ another construction of ${\widetilde K}$ is based
on the concept of boundary relations (see \cite{DHMS06}).
}
\end{rem}

A minimal self-adjoint dilation ${\widetilde K}$ for a scalar Sturm-Liouville
operator with a complex (dissipative) boundary condition has originally been constructed
by B.S. Pavlov in \cite{Pa1}. For the scalar case $(n=1)$ the operator in \eqref{SL-dilation} in the following 
example coincides with the one in \cite{Pa1}.

\begin{exam}\label{pavlov}
{\rm
Let $Q_+\in L^1_{\rm loc}(\dR_+,[\dC^n])$ be a matrix valued function such that $Q_+(\cdot)=Q_+(\cdot)^*$, and
let $A$ be the usual minimal operator in $\sH=L^2(\dR_+,\dC^n)$ associated to the Sturm-Liouville differential
expression $-\tfrac{d^2}{dx^2}+Q_+$,
\begin{equation*}
A=-\frac{d^2}{dx^2}+Q_+,\quad
\dom(A)=\bigl\{f\in\cD_{\rm max,+}:f(0)=f^\prime(0)=0\bigr\},
\end{equation*}
where $\cD_{\rm max,+}$ is the maximal domain defined by
\begin{displaymath}
\cD_{\rm max,+}=\bigl\{f\!\in\! L^2(\dR_+,\dC^n)\!:\!f,f^\prime\!\in\! AC(\dR_+,\dC^n),
-f^{\prime\prime}+Q_+f\in L^2(\dR_+,\dC^n) \bigr\}.
\end{displaymath}
It is well known that the adjoint operator $A^*$ is given by $A^*=-\tfrac{d^2}{dx^2}+Q_+$, $\dom (A^*)=\cD_{\rm max,+}$.

In the following we assume that the limit point case prevails at $+\infty$, so that the deficiency indices $n_\pm(A)$
of $A$ are both equal to $n$. In this case a boundary triplet $\Pi=\{\dC^n,\Gamma_0,\Gamma_1\}$ for $A^*$ is 
\begin{equation}\label{3.40A}
\Gamma_0 f:=f(0),\quad \Gamma_1 f:=f^\prime(0),\quad f\in\dom(A^*)=\cD_{\rm max,+}.
\end{equation}
For any  dissipative matrix $D\in[\dC^n]$ we consider the (maximal) dissipative
extension $A_D$ of $A$ determined by
       \begin{equation}\label{3.40}
A_D=A^*\upharpoonright\ker(\Gamma_1- D\Gamma_0),  \qquad \imag D\le 0.
     \end{equation}
\noindent
(a)  First suppose $0\in\rho(\imag D)$. Then ${\mathcal H}_D=\dC^n$ and by
Theorem  \ref{III.1} and Remark~\ref{specrem} the (minimal) self-adjoint dilation ${\widetilde K}$ of the
operator $A_D$ is a self-adjoint operator in $\sK=L^2(\dR_+,\dC^n) \oplus L^2(\dR,\dC^n)$
defined by
\begin{equation}\label{SL-dilation}
\begin{split}
\widetilde K&(f \oplus g)= \bigl(-f^{\prime\prime} + Q_+ f\bigr)\oplus -ig^\prime,\\
\dom(\widetilde K)& =
\left\{
\begin{matrix}f\in \cD_{\rm max,+},\, g\in W_2^1(\dR_-,\dC^n)\oplus W_2^1(\dR_+,\dC^n)\\
f'(0)-Df(0)=-i(-2 \imag D)^{1/2}g(0-),  \\
\  f'(0)-D^* f(0)=-i(-2 \imag D)^{1/2}g(0+)
\end{matrix}\right\}.
\end{split}
\end{equation}

\noindent
(b)  Let now $\ker(\imag D)\not =\{0\}$, so that  $\cH_D=\ran(\imag D)=\dC^k\not
=\dC^n$. According to Theorem \ref{III.1} the (minimal) self-adjoint
dilation ${\widetilde K}$ of the operator $A_D$ in $\sK=L^2(\dR_+,\dC^n) \oplus L^2(\dR,\dC^k)$  
is defined  by
\begin{equation*}
\begin{split}
\widetilde K&(f \oplus
 g)= \bigl(-f^{\prime\prime} + Q_+ f\bigr)\oplus -ig^\prime,\\
\dom(\widetilde K)& =
\left\{\begin{matrix} f\in \cD_{\rm max,+},\, g\in W_2^1(\dR_-,\dC^k)\oplus W_2^1(\dR_+,\dC^k)\\
P_D[f'(0)-Df(0)] =-i(-2 P_D\imag(D)\!\upharpoonright_{\cH_D})^{1/2}g(0-),  \\
P_D[f'(0)-D^* f(0)] = -i(-2 P_D\imag(D)\!\upharpoonright_{\cH_D})^{1/2}g(0+),\\
f'(0)- \Real(D)f(0)\in\cH_D
\end{matrix}\right\}.
\end{split}
\end{equation*}
}
\end{exam}

\subsection{Dilations and dissipative scattering systems}\la{dilatscat}

Let, as in the previous section, $A$ be a densely defined closed simple symmetric operator in
$\gotH$ with equal finite deficiency indices and let $\Pi=\{\cH,\Gamma_0,\Gamma_1\}$
be a boundary triplet for $A^*$, $A_0=A^*\upharpoonright\ker\Gamma_0$, with corresponding
Weyl function $M(\cdot)$. Let $D\in[\cH]$ be a dissipative matrix and let
$A_D=A^*\upharpoonright\ker(\Gamma_1-D\Gamma_0)$ be the corresponding maximal dissipative
extension in $\gotH$. Since $\dC_+\ni\lambda\mapsto M(\lambda)-D$ is a Nevanlinna function the limits
\begin{equation*}
M(\lambda+i0)-D=\lim_{\epsilon\rightarrow +0} M(\lambda+i0)-D
\end{equation*}
and
\begin{equation*}
N_D(\lambda+i0)=\lim_{\epsilon\rightarrow +0}N_D(\lambda+i\epsilon)
=\lim_{\epsilon\rightarrow +0}\bigl(D-M(\lambda+i\epsilon)\bigr)^{-1}
\end{equation*}
exist for a.e. $\lambda\in\dR$. We denote these sets of real points $\lambda$
by $\Sigma^M$ and $\Sigma^{N_D}$. Then we have
\begin{equation}\label{nd}
N_D(\lambda+i0)=\bigl(D-M(\lambda+i0)\bigr)^{-1},\qquad\lambda\in\Sigma^M\cap\Sigma^{N_D},
\end{equation}
cf. Section~\ref{scatsec}. Let $G$ be the symmetric first order differential operator in $L^2(\dR,\cH_D)$ and let
$\Pi_G=\{\cH_D,\gY_0,\gY_1\}$ be the boundary triplet from Lemma~\ref{disslem}.
Then $G_0=G^*\!\upharpoonright\ker(\gY_0)$ is the usual self-adjoint
differentiation operator in $L^2(\dR,\cH_D)$ and $K_0=A_0\oplus G_0$ is self-adjoint
in $\gotK=\gotH\oplus L^2(\dR,\cH_D)$. In the next theorem we consider the complete scattering system
$\{\widetilde K,K_0\}$, where $\widetilde K$ is the minimal
self-adjoint dilation of $A_D$ in $\gotK$ from Theorem~\ref{III.1}.

\begin{thm}\label{dilscat}
Let $A$, $\Pi=\{\cH,\Gamma_0,\Gamma_1\}$, $M(\cdot)$ and $A_D$ be as above and define $\cH_{M(\lambda)}$,
$\lambda\in\Sigma^M$, as in \eqref{hm}. Let $K_0=A_0\oplus G_0$ and let $\widetilde K$ be the minimal
self-adjoint dilation of $A_D$ from Theorem~{\rm \ref{III.1}}. Then the following holds.
\begin{itemize}
\item [{\rm (i)}] $K^{ac}_0=A^{ac}_0\oplus G_0$ is unitarily equivalent to the multiplication operator
with the free variable in $L^2(\bR,d\gl,\kH_{M(\gl)}\oplus\cH_D)$.
\item [{\rm (ii)}] In $L^2(\bR,d\gl,\kH_{M(\gl)}\oplus\cH_D)$
the scattering matrix $\{\widetilde S(\gl)\}$ of the complete scattering system $\{\widetilde K,K_0\}$ is given by
\begin{equation*}
\widetilde S(\gl) = \begin{pmatrix}I_{\kH_{M(\gl)}} & 0 \\ 0 & I_{\kH_D} \end{pmatrix} +2i\begin{pmatrix}
\widetilde T_{11}(\lambda) &  \widetilde T_{12}(\lambda)\\
 \widetilde T_{21}(\lambda) &  \widetilde T_{22}(\lambda)\end{pmatrix}\in[\cH_{M(\lambda)}\oplus\cH_D],
\end{equation*}
for all $\gl \in\Sigma^M\cap\Sigma^{N_D}$, where
\begin{equation*}
\begin{split}
\widetilde T_{11}(\lambda)&=P_{M(\lambda)}\sqrt{\imag(M(\lambda))}
                            \bigl(D - M(\gl)\bigr)^{-1}\sqrt{\imag(M(\lambda))}\upharpoonright_{\cH_{M(\lambda)}},\\
\widetilde T_{12}(\lambda)&=P_{M(\lambda)}\sqrt{\imag(M(\lambda))}
                            \bigl(D - M(\gl)\bigr)^{-1}\sqrt{-\imag(D)}\upharpoonright_{\cH_D},\\
\widetilde T_{21}(\lambda)&=P_D\sqrt{-\imag(D)}\bigl(D - M(\gl)\bigr)^{-1}\sqrt{\imag(M(\lambda))}\upharpoonright_{\cH_{M(\lambda)}},\\
\widetilde T_{22}(\lambda)&=P_D\sqrt{-\imag(D)}\bigl(D - M(\gl)\bigr)^{-1}\sqrt{-\imag(D)}
\upharpoonright_{\cH_D}\\
\end{split}
\end{equation*}
and $M(\lambda)=M(\lambda+i0)$.
\end{itemize}
\end{thm}

\begin{proof}
Let $K=A\oplus G$ and
let $\widetilde\Pi=\{\cH\oplus\cH_D,\widetilde\Gamma_0,\widetilde\Gamma_1\}$ be the boundary triplet
for $K^*$ from \eqref{3.12}. Notice that since $A$ and $G$ are densely defined closed simple
symmetric operators also $K$ is a densely defined closed simple symmetric operator.
Recall that for $\lambda\in\dC_+$ the Weyl function of
$\widetilde\Pi=\{\cH\oplus\cH_D,\widetilde\Gamma_0,\widetilde\Gamma_1\}$ is given by
\begin{equation}\label{wtmscat}
\widetilde M(\lambda)=\begin{pmatrix} M(\lambda)-\real(D) & 0 \\ 0 & -iP_D\imag (D)\upharpoonright_{\cH_D}\end{pmatrix}.
\end{equation}
Then Theorem~\ref{scattering} implies that
\begin{equation*}
L^2\bigl(\dR,d\lambda,\cH_{\widetilde M(\lambda)}\bigr),
\quad\cH_{\widetilde M(\lambda)}=\cH_{M(\lambda)}\oplus\cH_D,\quad\lambda\in\Sigma^M,
\end{equation*}
performs
a spectral representation of the absolutely continuous part
\begin{equation*}
\begin{split}
K_0^{ac}&=K_0\upharpoonright\dom(K_0)\cap\gotK^{ac}(K_0)\\
&=A_0\oplus G_0\upharpoonright\bigl(\dom(A_0)
\cap \gotH^{ac}(A_0)\bigr)\oplus L^2(\dR,\cH_D)=A^{ac}_0\oplus G_0
\end{split}
\end{equation*}
of $K_0$ such that the scattering matrix $\{\widetilde S(\lambda)\}$ of the scattering system $\{\widetilde K,K_0\}$
is given by
\begin{equation}\label{widescat}
\begin{split}
\widetilde S(\lambda)&=I_{\cH_{\widetilde M(\lambda)}}\\
&\qquad +2iP_{\widetilde M(\lambda)}\sqrt{\imag(\widetilde M(\lambda))}
\bigl(\widetilde\Theta-\widetilde M(\lambda)\bigr)^{-1}\sqrt{\imag(\widetilde M(\lambda))}
\upharpoonright_{\cH_{\widetilde M(\lambda)}}
\end{split}
\end{equation}
for all $\lambda\in\Sigma^{\widetilde M}\cap\Sigma^{N_{\widetilde\Theta}}$, where
$P_{\widetilde M(\lambda)}$ and $\upharpoonright_{\cH_{\widetilde M(\lambda)}}$ are the projection and
restriction in $\widetilde\cH=\cH\oplus\cH_D$ onto $\cH_{\widetilde M(\lambda)}$. Here $\widetilde\Theta$
is the self-adjoint relation from \eqref{wtdef} and the function $N_{\widetilde\Theta}$
is defined analogously to \eqref{ntheta} and
\begin{equation*}
N_{\widetilde\Theta}(\lambda+i0)=\bigl(\widetilde\Theta-\widetilde M(\lambda+i0)\bigr)^{-1}
\end{equation*}
holds for all
$\lambda\in\Sigma^{\widetilde M}\cap\Sigma^{N_{\widetilde\Theta}}$,
cf. \eqref{gvn}.

By \eqref{wtmscat} we have
\begin{equation*}
\sqrt{\imag(\widetilde M(\lambda+i0))}=\begin{pmatrix}\sqrt{\imag(M(\lambda+i0))} & 0 \\
0 & P_D \sqrt{-\imag(D)}\upharpoonright_{\cH_D}
\end{pmatrix}
\end{equation*}
for all $\lambda\in\Sigma^{\widetilde M}=\Sigma^M$ and \eqref{3.27} yields
\begin{equation*}
\bigl(\widetilde{\gT} - \widetilde{M}(\gl+i0)\bigr)^{-1} =
\begin{pmatrix}
(D - M(\gl+i0))^{-1} & (D - M(\gl+i0))^{-1}\upharpoonright_{\cH_D}\\
P_D(D - M(\gl+i0))^{-1} & P_D(D - M(\gl+i0))^{-1}\upharpoonright_{\cH_D}
\end{pmatrix}
\end{equation*}
for $\lambda\in\Sigma^M\cap\Sigma^{N_{\widetilde\Theta}}$. It follows that the sets
$\Sigma^M\cap\Sigma^{N_{\widetilde\Theta}}$ and $\Sigma^M\cap\Sigma^{N_D}$, see \eqref{nd}, coincide
and by inserting the above expressions into \eqref{widescat}
we conclude that for each $\lambda\in\Sigma^M\cap\Sigma^{N_D}$ the scattering matrix
$\{\widetilde S(\lambda)\}$ is a two-by-two block operator matrix
with respect to the decomposition
\begin{equation*}
\cH_{\widetilde M(\lambda)}=\cH_{M(\lambda)}\oplus\cH_D,\qquad\lambda\in\Sigma^M\cap\Sigma^{N_D},
\end{equation*}
with the entries from assertion (ii).
\end{proof}

\begin{rem}\label{invremdil}
{\rm
It is worth to note that the scattering matrix $\{\widetilde S(\lambda)\}$ of the scattering system
$\{\widetilde K,K_0\}$ in Theorem~\ref{dilscat} depends only on the dissipative matrix $D$ and the
Weyl function $M(\cdot)$ of the boundary triplet $\Pi=\{\cH,\Gamma_0,\Gamma_1\}$ for $A^*$.
In other words, the scattering matrix $\{\widetilde S(\lambda)\}$ is completely determined
by objects corresponding to the operators $A,A_0$ and $A_D$ in $\sH$.
}
\end{rem}

Let $A_D$ and $A_0$ be as in the beginning of this section.
In the following we will focus on the so-called
{\it dissipative scattering system} $\{A_D,A_0\}$ and we refer the reader to
\cite{D3,D2,Ma1,Na1,Na2,N81,N84,N85,N86} for a detailed investigation of such scattering systems.
We recall only that the wave operators $W_\pm(A_D,A_0)$ of the dissipative scattering
system $\{A_D,A_0\}$ are defined by
\begin{equation*}
W_+(A_D,A_0)=\slim_{t\rightarrow +\infty} e^{it A_D^*} e^{-itA_0}P^{ac}(A_0)
\end{equation*}
and
\begin{equation*}
W_-(A_D,A_0)=\slim_{t\rightarrow +\infty} e^{-it A_D} e^{itA_0}P^{ac}(A_0),
\end{equation*}
where $e^{-itA_D}:=\slim_{n\rightarrow\infty}(1+\tfrac{it}{n}A_D)^{-n}$, see e.g. \cite[$\S$IX]{Ka1}.
The scattering operator
\begin{equation*}
S_D:=W_+(A_D,A_0)^*W_-(A_D,A_0)
\end{equation*}
of the dissipative scattering system $\{A_D,A_0\}$ will be regarded as an operator in $\gotH^{ac}(A_0)$.
Then $S_D$
is a contraction
which in general is not unitary. Since $S_D$ and $A_0^{ac}$ commute it follows that $S_D$ is unitarily
equivalent to a multiplication operator induced by a family $\{S_D(\lambda)\}$ of contractive operators
in a spectral representation of $A_0^{ac}$.

With the help of Theorem~\ref{dilscat}
we obtain a representation of the scattering matrix of
the dissipative scattering system $\{A_D,A_0\}$ in terms of the Weyl function $M(\cdot)$
of $\Pi=\{\cH,\Gamma_0,\Gamma_1\}$ in the following corollary, cf. Theorem~\ref{scattering}.

\begin{cor}\label{disscatcor}
Let $A$, $\Pi=\{\cH,\Gamma_0,\Gamma_1\}$, $A_0=A^*\!\upharpoonright\ker(\Gamma_0)$, $M(\cdot)$ and $A_D$ be as above and define $\cH_{M(\lambda)}$,
$\lambda\in\Sigma^M$, as in \eqref{hm}. Then the following holds.
\begin{itemize}
\item [{\rm (i)}] $A^{ac}_0$ is unitarily equivalent to the multiplication operator
with the free variable in $L^2(\bR,d\gl,\kH_{M(\gl)})$.
\item [{\rm (ii)}] In $L^2(\bR,d\gl,\kH_{M(\gl)})$
the scattering matrix $\{S_D(\gl)\}$ of the dissipative scattering system $\{A_D,A_0\}$ is given by
{\small
\begin{displaymath}
S_D(\gl) =I_{\kH_{M(\gl)}}+2i P_{M(\lambda)}\sqrt{\imag(M(\lambda))}\bigl(D - M(\gl)\bigr)^{-1}\sqrt{\imag(M(\lambda))}
\upharpoonright_{\cH_{M(\lambda)}}
\end{displaymath}
}
for all $\lambda\in\Sigma^M\cap\Sigma^{N_D}$, where $M(\lambda)=M(\lambda+i0)$.
\end{itemize}
\end{cor}

\begin{proof}
Let $\widetilde K$ be the minimal self-adjoint dilation of $A_D$ from Theorem~\ref{III.1}. Since
for $t\geq 0$ we have
\begin{equation*}
P_\gotH e^{-it\widetilde K}\!\upharpoonright\gotH=\slim_{n\rightarrow\infty}
P_\gotH \bigl(1+\tfrac{it}{n}\widetilde K\bigr)^{-n}
\!\upharpoonright_\gotH=\slim_{n\rightarrow\infty} \bigl(1+\tfrac{it}{n}A_D\bigr)^{-n}=
e^{-itA_D}
\end{equation*}
it follows that the wave operators $W_+(A_D,A_0)$ and $W_-(A_D,A_0)$ coincide with
\begin{equation*}
\begin{split}
P_\gotH W_+(\widetilde K,K_0)\upharpoonright_{\gotH}&=\slim_{t\rightarrow +\infty} P_\gotH e^{it\widetilde K}
e^{-itK_0}P^{ac}(K_0)\upharpoonright_\gotH\\
&=\slim_{t\rightarrow +\infty} P_\gotH e^{it\widetilde K} \upharpoonright_{\gotH}
e^{-itA_0}P^{ac}(A_0)
\end{split}
\end{equation*}
and
\begin{equation*}
\begin{split}
P_\gotH W_-(\widetilde K,K_0)\upharpoonright_{\gotH}&=\slim_{t\rightarrow -\infty} P_\gotH e^{it\widetilde K}
e^{-itK_0}P^{ac}(K_0)\upharpoonright_\gotH\\
&=\slim_{t\rightarrow +\infty} P_\gotH e^{-it\widetilde K} \upharpoonright_{\gotH}
e^{itA_0}P^{ac}(A_0),
\end{split}
\end{equation*}
respectively.
This implies that the scattering operator $S_D$ coincides with the compression $P_{\gotH^{ac}(A_0)}
S(\widetilde K,K_0)\upharpoonright_{\gotH^{ac}(A_0)}$ of the scattering operator $S(\widetilde K,K_0)$
onto $\gotH^{ac}(A_0)$. Therefore
the scattering matrix $S_D(\lambda)$ of the dissipative
scattering system is given by the upper left corner
\begin{equation*}
\bigl\{I_{\cH_{M(\lambda)}}+2i\widetilde T_{11}(\lambda)\bigr\},\quad\lambda\in\Sigma^M\cap\Sigma^{N_D},
\end{equation*}
of
the scattering matrix $\{\widetilde S(\lambda)\}$ of the scattering
system $\{\widetilde K,K_0\}$, see Theorem~\ref{dilscat}.
\end{proof}

\subsection{Lax-Phillips scattering systems}\label{laxsubsec}

Let again $A$, $\Pi=\{\cH,\Gamma_0,\Gamma_1\}$, $\{A_D,A_0\}$ and $G$, $G_0$,
$\Pi_G=\{\cH_D,\gY_0,\gY_1\}$ be as in the previous subsections. In
Corollary~\ref{disscatcor} we have shown that the scattering matrix of the dissipative
scattering system $\{A_D,A_0\}$ is the left upper corner in the block operator matrix
representation of the scattering matrix $\{\widetilde S(\lambda)\}$ of the scattering
system $\{\widetilde K,K_0\}$, where $\widetilde K$ is a minimal self-adjoint dilation of
$A_D$ in $\gotK = \gotH\oplus L^2(\dR,\cH_D)$ and $K_0=A_0\oplus G_0$, cf.
Theorem~\ref{dilscat}.

In the following we are going to interpret the right lower corner
of $\{\widetilde S(\lambda)\}$ as the scattering matrix corresponding
to a Lax-Phillips scattering system, see e.g. \cite{BW,LP} for further details.
To this end we decompose the space
$L^2(\dR,\cH_D)$ into the orthogonal sum of the subspaces
\begin{equation}\label{dpm}
\kD_-:=L^2(\dR_-,\cH_D)\quad\text{and}\quad\kD_+:=L^2(\dR_+,\cH_D).
\end{equation}
Then clearly $\gotK=\gotH\oplus\cD_-\oplus\cD_+$ and we agree to denote the elements
in $\gotK$ in the form $f\oplus g_-\oplus g_+$, $f\in\gotH$, $g_\pm\in\cD_\pm$ and
$g=g_-\oplus g_+\in L^2(\dR,\cH_D)$.
By $J_+$ and $J_-$ we denote the operators
\begin{equation*}
J_+: L^2(\dR,\cH_D)\rightarrow\gotK,\quad g\mapsto 0\oplus 0\oplus g_+,
\end{equation*}
and
\begin{equation*}
J_-: L^2(\dR,\cH_D)\rightarrow\gotK,\quad g\mapsto 0\oplus g_-\oplus 0,
\end{equation*}
respectively.
Notice that $J_++J_-$ is the embedding of $L^2(\dR,\cH_D)$ into $\gotK$.
In the next lemma we show that $\cD_+$ and $\cD_-$ are so-called
{\it outgoing} and {\it incoming subspaces} for the self-adjoint dilation
$\widetilde K$ in $\gotK$.
\begin{lem}\label{III.3}
Let $\widetilde K$ be the self-adjoint operator from Theorem~{\rm \ref{III.1}}, let
$\cD_\pm$ be as in \eqref{dpm} and $A_0=A^*\upharpoonright\ker(\Gamma_0)$ be as above.
Then 
        \begin{equation*}
e^{-it\widetilde K} \subseteq\cD_\pm,\,\,\,\,t\in\dR_\pm,  \quad\text{and}\quad
\bigcap_{t\in \bR}e^{-it\widetilde K}\kD_\pm = \{0\},
       \end{equation*}
and, if in addition $\sigma(A_0)$ is singular, then
\begin{equation}\label{3.27x}
\overline{\bigcup_{t\in \bR}e^{-it\widetilde K}\kD_+}= \overline{\bigcup_{t\in
\bR}e^{-it\widetilde K}\kD_-}=\gotK^{ac}(\widetilde{K}).
\end{equation}
\end{lem}

\begin{proof}
Let us first show that
\begin{equation}\la{3.39}
e^{-it\widetilde K}\upharpoonright\kD_\pm =
J_\pm e^{-itG_0}\upharpoonright\kD_\pm, \qquad t \in \bR_\pm,
\end{equation}
holds. In fact,
since $e^{-itG_0}$ is the right shift group we
have
\begin{equation*}
e^{-itG_0}(\dom(G) \cap \kD_\pm) \subseteq  \dom(G) \cap
\kD_\pm, \quad t \in \bR_\pm,
\end{equation*}
where $\dom(G) \cap \kD_\pm = \{W^{1,2}(\bR,\kH_D):
f(x) = 0,\, x \in \bR_\pm\}$.
Let us fix some $t\in\dR_\pm$ and denote the symmetric operator $A\oplus G$
by $K$.
Since
\begin{equation*}
J_\pm\bigl(\dom(G) \cap \kD_\pm\bigr) \subset \dom(K) \subset
\dom(\widetilde K)
\end{equation*}
the function
\begin{equation*}
f_{t,\pm}(s) :=
e^{i(s-t)\widetilde K}J_\pm e^{-isG_0}\upharpoonright_{\cD_\pm} f_\pm,\quad s \in \bR_\pm,\quad f_\pm\in
\dom(G) \cap \kD_\pm,
\end{equation*}
is differentiable and
\begin{equation*}
\frac{d}{ds}f_{t,\pm}(s) =
ie^{i(s-t)\widetilde K}\bigl(\widetilde K - 0_\gotH\oplus G_0\bigr)J_\pm e^{-isG_0}\upharpoonright_{\cD_\pm}f_\pm =
0, \quad t \in \bR_\pm,
\end{equation*}
holds. Hence we have $f_{t,\pm}(0)=f_{t,\pm}(t)$ and together with the observation that
the set $\dom(G) \cap \kD_\pm$ is dense in $\kD_\pm$
this immediately implies \eqref{3.39}. Then we
obtain
$e^{-it\widetilde K}\kD_\pm \subseteq \kD_\pm$, $t \in
\bR_\pm$ and
\begin{equation*}
\bigcap_{t\in \bR}e^{-it\widetilde K}\kD_\pm \subseteq
\bigcap_{t\in \bR_\pm}e^{-it\widetilde K}\kD_\pm =
\bigcap_{t\in \bR_\pm}J_\pm e^{-itG_0}\kD_\pm =\{0\}.
\end{equation*}
Let us show \eqref{3.27x}.
Since $A$ has finite deficiency indices the wave operators
$W_\pm(\widetilde{K},A_0 \oplus G_0)$ exist and are complete, i.e.,
$\ran(W_\pm(\widetilde{K},A_0 \oplus G_0)) =\gotK^{ac}(\widetilde{K})$ holds.
Since $A_0$ is singular we have
\bed
W_\pm(\widetilde{K},A_0 \oplus G_0) =
\slim_{t\to\pm\infty}e^{it\widetilde{K}}(J_++J_-)e^{-itG_0}\!\upharpoonright_{L^2}
\eed
and it follows from \eqref{3.39} that $W_\pm(\widetilde{K},A_0 \oplus G_0)f_\pm=f_\pm$ for $f_\pm\in\cD_\pm$,
so that in particular $\cD_\pm$ and $e^{-itG_0}\cD_\pm\in\gotK^{ac}(\widetilde{K})$ for $t\in\dR_\pm$.
Assume now that
$g\in L^2(\dR,\cH_D)$ vanishes identically on some open interval $(-\infty,\alpha)$. Then for $r>0$ sufficiently
large $e^{-irG_0}g\in\cD_+$ and by \eqref{3.39} for $t>r$
\begin{equation*}
e^{it\widetilde K}(J_++J_-)e^{-i(t-r)G_0}e^{-irG_0}g=e^{ir\widetilde K}J_+e^{-irG_0}g.
\end{equation*}
Since the elements $g\in L^2(\dR,\cH_D)$ which vanish on intervals $(-\infty,\alpha)$ form a dense set
in $L^2(\dR,\cH_D)$ and the wave operator $W_+(\widetilde{K},A_0 \oplus G_0)$ is complete we conclude that
\begin{equation}\label{asx}
\bigcup_{r\in\dR_+}e^{ir\widetilde K}\cD_+
\end{equation}
is a dense set in $\sK^{ac}(\widetilde K)$. A similar argument shows that the set \eqref{asx} with $\dR_+$ and
$\cD_+$ replaced by $\dR_-$ and $\cD_-$, respectively, is also dense in $\sK^{ac}(\widetilde K)$. This implies \eqref{3.27x}.
\end{proof}
According to Lemma \ref{III.3} the system $\{\widetilde K,\cD_-,\cD_+\}$ is a Lax-Phillips scattering system
and in particular
the {\it Lax-Phillips wave operators}
\begin{equation*}
\gO_\pm := \slim_{t\to\pm\infty}e^{it\widetilde K}J_\pm
e^{-itG_0}:L^2(\dR,\cH_D)\rightarrow\gotK
\end{equation*}
exist, cf. \cite{BW}.
We note that $\slim_{t\to\pm\infty}J_\mp e^{-itG_0} = 0$ and therefore
the restrictions of the wave operators $W_\pm(\widetilde K,K_0)$
of the scattering system $\{\widetilde K,K_0\}$, $K_0=A_0\oplus G_0$, onto $L^2(\dR,\cH_D)$,
\begin{equation*}
W_\pm(\widetilde K,K_0)\upharpoonright_{L^2} =
\slim_{t\to\pm\infty}e^{it\widetilde K}(J_++J_-) e^{-itG_0},
\end{equation*}
coincide with the Lax-Phillips wave operators $\Omega_\pm$.
Hence the {\it Lax-Phillips scattering operator} $S^{LP} := \gO^*_+\gO_-$
admits the representation
\begin{equation*}
S^{LP}= P_{L^2}S(\widetilde K,K_0)\upharpoonright_{L^2}
\end{equation*}
where $S(\widetilde K,K_0)= W_+(\widetilde K,K_0)^*W_-(\widetilde K,K_0)$ is the scattering operator
of the scattering system $\{\widetilde K,K_0\}$. The Lax-Phillips scattering operator $S^{LP}$
is a contraction in $L^2(\dR,\cH_D)$ and commutes with the self-adjoint differential operator $G_0$.
Hence $S^{LP}$ is unitarily equivalent to a multiplication operator
induced by a family $\{S^{LP}(\lambda)\}$ of contractive operators in $L^2(\dR,\cH_D)$, this family
is called the {\it Lax-Phillips scattering matrix}.

The above considerations together with Theorem~\ref{dilscat} immediately
imply the following corollary on the representation of the Lax-Phillips scattering matrix.

\begin{cor}\label{lax1}
Let $\{\widetilde K,\cD_-,\cD_+\}$ be the Lax-Phillips scattering
system considered in Lemma~{\rm \ref{III.3}} and
let $A$, $\Pi=\{\cH,\Gamma_0,\Gamma_1\}$, $A_D$, $M(\cdot)$ and $G_0$ be as
in the previous subsections. Then $G_0 =G_0^{ac}$ is unitarily
equivalent to the multiplication operator
with the free variable in
$L^2(\dR,\cH_D)=L^2(\dR,d\lambda,\kH_D)$ and the Lax-Phillips scattering matrix
$\{S^{LP}(\gl)\}$ admits the representation
\begin{equation}\la{5.11}
S^{LP}(\lambda)= I_{\kH_D}+2iP_D\sqrt{\imag(-D)}\bigl(D-M(\lambda)\bigr)^{-1}
\sqrt{\imag(-D)}\upharpoonright_{\cH_D}
\end{equation}
for $\lambda\in\Sigma^M\cap\Sigma^{N_D}$, where $M(\lambda)=M(\lambda+i0)$.
\end{cor}

Let again $A_D$ be the maximal dissipative extension of $A$ corresponding to
the maximal dissipative matrix $D\in[\cH]$ and let $\cH_D=\ran(\imag (D))$. By
\cite{DM92} the characteristic function $W_{A_D}$ of the completely non-self-adjoint part of
$A_D$ is given by
\begin{equation}\label{2.81}
\begin{split}
W_{A_D} :\dC_-&\rightarrow [\cH_D]\\
\mu\mapsto I_{\kH_D}&
-2iP_D\sqrt{-\imag(D)}\bigl(D^* - M(\mu)\bigr)^{-1}\sqrt{-\imag(D)}\upharpoonright_{\cH_D}.
\end{split}
\end{equation}

Comparing \eqref{5.11} and \eqref{2.81} we obtain the famous relation
between the Lax-Phillips scattering matrix and the characteristic function found by
Adamyan and Arov in \cite{AA1,AA2,AA3,AA4}.

\begin{cor}\label{adamyanarov}
Let the assumption be as in Corollary~{\rm \ref{lax1}}. Then the Lax-Phillips scattering matrix
$\{S^{LP}(\lambda)\}$ and the characteristic function $W_{A_D}$ of the
maximal dissipative operator $A_D$ are related by
\begin{equation*}
S^{LP}(\gl) = W_{A_D}(\gl - i0)^*,\qquad \lambda\in\Sigma^M\cap\Sigma^{N_D}.
\end{equation*}
\end{cor}

Next we consider the special case that the spectrum $\sigma(A_0)$ of the self-adjoint extension
$A_0=A^*\upharpoonright\ker(\Gamma_0)$ is purely singular, $\gotH^{ac}(A_0)=\{0\}$. As usual
let $M(\cdot)$ be the Weyl function corresponding to $\Pi=\{\cH,\Gamma_0,\Gamma_1\}$.
Then we have $\cH_{M(\lambda)}=\ran(\imag(M(\lambda+i0)))=\{0\}$ for a.e. $\lambda\in\Sigma^M$,
cf. \cite{BMN1}, and if even $\sigma(A_0)=\sigma_p(A_0)$ then $\cH_{M(\lambda)}=\{0\}$ for
all $\lambda\in\Sigma^M$.
Therefore
Theorem~\ref{dilscat} and Corollaries~\ref{lax1} and~\ref{adamyanarov}
imply the following statement.

\begin{cor}\label{singlax}
Let the assumption be as in Corollary~{\rm \ref{lax1}}, let $K_0=A_0\oplus G_0$ and
assume in addition that $\sigma(A_0)$ is purely singular. Then
the scattering matrix
$\{\widetilde S(\lambda)\}$ of the complete scattering system $\{\widetilde K,K_0\}$
coincides with the Lax-Phillips scattering matrix
$\{S^{LP}(\lambda)\}$ of the Lax-Phillips scattering system $\{\widetilde K,\cD_-,\cD_+\}$,
that is,
\begin{equation}\label{lpchar}
\widetilde S(\lambda)=S^{LP}(\lambda)= W_{A_D}(\gl - i0)^*
\end{equation}
for a.e. $\lambda\in\dR$. If even $\sigma(A_0)=\sigma_p(A_0)$, then \eqref{lpchar}
holds for all $\lambda\in\Sigma^M\cap\Sigma^{N_D}$.
\end{cor}

\subsection{A dissipative Schr\"{o}dinger-Poisson system}\label{III}

In this subsection we consider an open quantum system consisting of a
self-adjoint and a maximal dissipative extension of a symmetric regular Sturm-Liouville
differential operator. Such maximal dissipative operators or pseudo-Hamiltonians are used
in the description of carrier transport in semi-conductors, see e.g.
\cite{baro1,BKNR1,wr2,wr1,KNR1,KN1,KL}.

Assume that
$-\infty<x_l<x_r<\infty$ and let $V\in L^\infty((x_l,x_r))$ be a real valued function. Moreover let
$m\in L^\infty((x_l,x_r))$ be a real function such that $m>0$ and $m^{-1}\in L^\infty((x_l,x_r))$.
It is well-known that
\begin{equation*}
\begin{split}
(Af)(x) &:=
-\frac{1}{2}\frac{d}{dx}\frac{1}{m(x)}\frac{d}{dx}f(x) + V(x)f(x),\\
\dom(A) & :=  \left\{f \in L^2((x_l,x_r)): \ba{l}
f, \frac{1}{m}f' \in W^1_2((x_l,x_r)) \\
f(x_l) = f(x_r)= 0\\
\left(\frac{1}{m}f'\right)(x_l) = \left(\frac{1}{m}f'\right)(x_r) = 0
\ea
\right\},
\end{split}
\end{equation*}
is a densely defined closed simple symmetric operator in the Hilbert space $\gotH := L^2((x_l,x_r))$.
The deficiency indices of $A$ are $n_+(A) = n_-(A) = 2$ and the adjoint
operator $A^*$ is given by
\begin{equation*}
\begin{split}
(A^*f)(x) & =  -\frac{1}{2}\frac{d}{dx}\frac{1}{m(x)}\frac{d}{dx}f(x) + V(x)f(x),\\
\dom(A^*) & =  \left\{f \in \gotH: f,\frac{1}{m}f' \in
W^1_2((x_l,x_r))\right\}.
\end{split}
\end{equation*}
It is straightforward to verify that $\Pi=\{\bC^2,\gG_0,\gG_1\}$, where
\begin{equation}\label{2.13}
\gG_0f := \left(
\begin{array}{c}
f(x_l)\\
f(x_r)
\end{array}
\right)
\quad \mbox{and} \quad
\gG_1f := \left(
\begin{array}{c}
\left(\frac{1}{2m}f'\right)(x_l)\\
-\left(\frac{1}{2m}f'\right)(x_r)
\end{array}
\right),
\end{equation}
$f\in \dom(A^*)$, is a boundary triplet for $A^*$.
Notice that the self-adjoint extension
$A_0 = A^*\!\upharpoonright\ker(\gG_0)$ corresponds to Dirichlet
boundary conditions, that is,
\begin{equation*}
\dom(A_0) = \left\{f \in \gotH: f,\frac{1}{m}f' \in
W^1_2((x_l,x_r)), f(x_l) = f(x_r) = 0 \right\}.
\end{equation*}
It is well known that $A_0$ is semibounded from below and that $\sigma(A_0)$ consists
of eigenvalues accumulating to $+\infty$.
As usual we denote the Weyl function corresponding to $\Pi=\{\dC^2,\Gamma_0,\Gamma_1\}$
by $M(\cdot)$. Here $M(\cdot)$
is a two-by-two matrix-valued function which has poles at the eigenvalues of $A_0$
and in particular we have
\begin{equation}\label{hma0}
\cH_{M(\lambda)}=\ran\bigl(\imag(M(\lambda))\bigr)=\{0\}\quad\text{for all}\quad \lambda\in\Sigma^M.
\end{equation}
If $\varphi_\lambda,\psi_\lambda\in L^2((x_l,x_r))$ are fundamental solutions of
$-\tfrac{1}{2}\bigl(\frac{1}{m}f^\prime\bigr)^\prime+Vf=\lambda f$ satisfying the boundary conditions
\begin{equation}\label{fundsol}
\varphi_\lambda(x_l)=1,\,\,\,\,\bigl(\tfrac{1}{m}\varphi_\lambda^\prime\bigr)(x_l)=0,\qquad
\psi_\lambda(x_l)=0,\,\,\,\,\bigl(\tfrac{1}{m}\psi_\lambda^\prime\bigr)(x_l)=1,
\end{equation}
then $M$ can be written as
\begin{equation}\label{fsm}
M(\lambda)=\frac{1}{2\psi_\lambda(x_r)}\begin{pmatrix} -\varphi_\lambda(x_r) & 1\\
1 & -\bigl(\tfrac{1}{m}\psi_\lambda^\prime\bigr)(x_r)\end{pmatrix},\qquad \lambda\in\rho(A_0).
\end{equation}
We are interested in  maximal dissipative extensions
\begin{equation*}
A_D=A^*\upharpoonright\ker(\Gamma_1-D\Gamma_0)
\end{equation*}
of $A$ where $D\in[\dC^2]$ has the special form
\begin{equation}\label{dspec}
D = \begin{pmatrix}
-\gk_l & 0\\
0 & -\gk_r
\end{pmatrix},
\quad \imag(\gk_l) \ge 0, \quad \imag(\gk_r) \ge 0.
\end{equation}
Of course, if both $\kappa_l$ and $\kappa_r$ are real constants then $\cH_D=\ran(\imag(D))=\{0\}$
and $A_D$ is self-adjoint. In this case $A_D$ can be identified with the self-adjoint dilation
$\widetilde K$ acting in $\gotH\oplus L^2(\dR,\{0\})\widetilde=\gotH$, cf. Theorem~\ref{III.1}.

Let us first consider the situation where both $\kappa_l$ and $\kappa_r$ have positive
imaginary parts. Then $\cH_D=\dC^2$ and the self-adjoint dilation $\widetilde K$
from Theorem~\ref{III.1} is given by
\begin{equation*}
\begin{split}
\widetilde K&(f\oplus g_-\oplus g_+)=\bigl(-\tfrac{1}{2}\bigl(\tfrac{1}{m}f^{\prime}\bigr)^\prime
+Vf\bigr)\oplus -ig_-^\prime\oplus -ig_+^\prime,\\
\dom\widetilde K&=\left\{\begin{matrix}f,\tfrac{1}{m}f^\prime\in W^1_2((x_l,x_r)),\\
g_\pm\in W^1_2(\dR_\pm,\dC^2)\end{matrix}:
\begin{matrix}
\Gamma_0f-\gY_0 g=0,\\
(\Gamma_1-\real(D)\Gamma_0)f+\gY_1 g=0
\end{matrix}\right\}.
\end{split}
\end{equation*}
Here $\Pi_G=\{\dC^2,\gY_0,\gY_1\}$ is the boundary triplet for first order differential operator $G\subset G^*$
in $L^2(\dR,\dC^2)$ from Lemma~\ref{disslem} and
we have decomposed the elements $f\oplus g$ in $\gotH\oplus L^2(\dR,\dC^2)$
as agreed in the beginning of Section~\ref{laxsubsec}. Let us set
\begin{equation*}
g_-(0-)=\begin{pmatrix} g_l(0-)\\ g_r(0-)\end{pmatrix}\quad\text{and}\quad
g_+(0+)=\begin{pmatrix}g_l(0+)\\ g_r(0+)\end{pmatrix}.
\end{equation*}
Then a straightforward calculation
using the definitions of $\Pi=\{\dC^2,\Gamma_0,\Gamma_1\}$ and $\Pi_G=\{\dC^2,\gY_0,\gY_1\}$
in \eqref{2.13} and Lemma~\ref{disslem}, respectively, shows that an element $f\oplus g_-\oplus g_+$
belongs to $\dom(\widetilde K)$ if and only if
\begin{equation*}
\begin{split}
\bigl(\tfrac{1}{2m}f'\bigr)(x_l) + \gk_lf(x_l) & =  -i \sqrt{2\imag(\kappa_l)}g_l(0-) \\
\bigl(\tfrac{1}{2m}f'\bigr)(x_l) + \overline{\gk}_lf(x_l)  & =  -i\sqrt{2\imag(\kappa_l)}g_l(0+)\\
\bigl(\tfrac{1}{2m}f'\bigr)(x_r) - \gk_rf(x_r) & =  i\sqrt{2\imag(\kappa_r)}g_r(0-) \\
\bigl(\tfrac{1}{2m}f'\bigr)(x_r) - \overline{\gk}_rf(x_r)  & = i\sqrt{2\imag(\kappa_r)} g_r(0+)
\end{split}
\end{equation*}
holds. We note that this dilation $\widetilde K$ is isomorph in the
sense of \cite[Section I.4]{FN} to those
used in
\cite{BKNR1,BKNR2,KNR1,KN1}.

Theorem~\ref{dilscat} and the fact that $\sigma(A_0)$ is singular (cf. \eqref{hma0}) imply that
the scattering matrix $\{\widetilde S(\lambda)\}$ of the scattering system
$\{\widetilde K,K_0\}$, $K_0=A_0\oplus G_0$,
coincides with
\begin{equation*}
S^{LP}(\lambda)=I_{\dC^2}+2i
\sqrt{-\imag(D)}\bigl(D-M(\lambda)\bigr)^{-1}
\sqrt{-\imag(D)}\in[\dC^2]
\end{equation*}
for all $\lambda\not\in\sigma_p(A_0)\cap\dR$, where $M(\lambda)=M(\lambda+i0)$ (cf. Corollary~\ref{singlax}).
By \eqref{dspec} here
$\sqrt{-\imag(D)}$ is a diagonal matrix with entries $\sqrt{\imag(\kappa_l)}$ and $\sqrt{\imag(\kappa_r)}$.
We leave it to the reader to compute $S^{LP}(\lambda)$ explicitely in terms of the fundamental solutions
$\varphi_\lambda$ and $\psi_\lambda$ in \eqref{fundsol}.
According to Corollary~\ref{adamyanarov} the continuation of the characteristic function $W_{A_D}$ of the completely
non-self-adjoint pseudo-Hamiltonian $A_D$ from $\dC_-$ to $\dR\backslash\{\sigma_p(A_0)\}$
coincides with $S^{LP}(\lambda)^*$,
\begin{equation*}
W_{A_D}(\lambda-i0)=I_{\dC^2}-2i
\sqrt{-\imag(D)}\bigl(D^*-M(\overline\lambda)\bigr)^{-1}
\sqrt{-\imag(D)}=S^{LP}(\lambda)^*.
\end{equation*}

Next we consider briefly the case where one of the entries of $D$ in \eqref{dspec} is real.
Assume e.g. $\kappa_l\in\dR$. In this case $\cH_D=\dC\widetilde=\{0\}\oplus\dC$, $P_D$ is
the orthogonal projection onto the second component in $\dC^2$
and $G$ is a first order differential operator in $L^2(\dR,\dC)$.
The self-adjoint dilation $\widetilde K$ is
\begin{equation*}
\begin{split}
\widetilde K&(f\oplus g_-\oplus g_+)=\bigl(-\tfrac{1}{2}\bigl(\tfrac{1}{m}f^{\prime}\bigr)^\prime
+Vf\bigr)\oplus -ig_-^\prime\oplus -ig_+^\prime,\\
\dom\widetilde K&=\left\{\begin{matrix}f,\tfrac{1}{m}f^\prime\in W^1_2((x_l,x_r)),\\
g_\pm\in W^1_2(\dR_\pm,\dC^2)\end{matrix}:
\begin{matrix}
P_D\Gamma_0f-\gY_0 g=0,\\
(1-P_D)(\Gamma_1-\real(D)\Gamma_0)f=0,\\
P_D(\Gamma_1-\real(D)\Gamma_0)f+\gY_1 g=0
\end{matrix}\right\},
\end{split}
\end{equation*}
and explicitely this means that an element $f\oplus g_-\oplus g_+$ belongs to $\dom(\widetilde K)$
if and only if
\begin{equation*}
\begin{split}
\bigl(\tfrac{1}{2m}f\bigr)^\prime(x_r)-\overline\kappa_rf(x_r)&=i\sqrt{2\imag(\kappa_r)}g_+(0+)\\
\bigl(\tfrac{1}{2m}f\bigr)^\prime(x_r)-\kappa_rf(x_r)&=i\sqrt{2\imag(\kappa_r)}g_-(0-)\\
\bigl(\tfrac{1}{2m}f\bigr)^\prime(x_l)+\kappa_lf(x_l)&=0
\end{split}
\end{equation*}
holds. The scattering matrix of $\{\widetilde K,K_0\}$ is given by
\begin{equation*}
S^{LP}(\lambda)=I_{\cH_D}+2i \imag(\kappa_r)P_D\bigl(D-M(\lambda)\bigr)^{-1}\upharpoonright_{\cH_D},\quad\lambda\in\Sigma^M,
\end{equation*}
which is now a scalar function, and is related to the characteristic function of the maximal dissipative operator
$A_D$ by $S^{LP}(\lambda)=W_{A_D}(\lambda-i0)^*$.

\section{Energy dependent scattering systems}\la{couplescat}

In this section we consider families $\{A_{-\tau(\lambda)},A_0\}$ of
scattering systems, where $\tau(\cdot)$ is a matrix Nevanlinna
function and $\{A_{-\tau(\lambda)}\}$ is a family of maximal
dissipative extensions of a symmetric operator $A$ with finite
deficiency indices.  Such scattering systems arise naturally in the
description of open quantum systems, see e.g. Section~\ref{exam2}
where a simple model of a so-called quantum transmitting Schr\"{o}dinger-Poisson system
is described. Following ideas in \cite{DHMS00} (see also \cite{BL07,CDR01,DL95,HKS97,HKS98})
the family $\{A_{-\tau(\lambda)}\}$
is ``linearized'' in an abstract way, that is, we construct a self-adjoint extension
$\widetilde L$ of $A$ which acts in a larger Hilbert space $\sH\oplus\sG$
and satisfies
\begin{equation*}
P_\sH\bigl(\widetilde L-\lambda\bigr)^{-1}\!\upharpoonright_\sH=\bigl(A_{-\tau(\lambda)}-\lambda\bigr)^{-1},
\end{equation*}
so that, roughly speaking, the
open quantum system is embedded into a closed system. The
corresponding Hamiltonian $\widetilde L$ is semibounded if and only if
$A_0$ is semibounded and $\tau(\cdot)$ is holomorphic on some interval
$(-\infty,\eta)$. The essential observation here is that the
scattering matrix of $\{\widetilde L,L_0\}$, where $L_0$ is the direct
orthogonal sum of $A_0$ and a self-adjoint operator connected with
$\tau(\cdot)$, pointwise coincides with the scattering matrix of a
scattering system $\{\widetilde K,K_0\}$ as investigated in the
previous section. From a physical point of view this in particular
justifies the use of quasi-Hamiltonians $\widetilde K$ for the
analysis of scattering processes in suitable small energy ranges.

\subsection{The \v{S}traus family and its characteristic functions}\label{straussf}

Let $A$ be a densely defined closed simple symmetric operator in the separable
Hilbert space $\gotH$ with equal finite deficiency indices $n_\pm(A)=n<\infty$ and
let $\Pi=\{\kH,\gG_0,\gG_1\}$ be a boundary triplet for $A^*$.
Assume that $\tau(\cdot)$ is an $[\cH]$-valued Nevanlinna function
and consider the family $\{A_{-\tau(\lambda)}\}$,
\begin{equation*}
A_{-\tau(\lambda)}:=A^*\upharpoonright\ker\bigl(\Gamma_1+\tau(\lambda)\Gamma_0\bigr),\qquad\lambda\in\dC_+,
\end{equation*}
of closed extension of $A$. Sometimes it is convenient to consider $A_{-\tau(\lambda)}$
for all $\lambda\in\mathfrak h(\tau)$, that is, for all $\lambda\in\dC\backslash\dR$ and all real
points $\lambda$ where $\tau$ is holomorphic, cf. Section~\ref{weylsec}.
Since $\imag\tau(\lambda)\geq 0$ for $\lambda\in\dC_+$ it
follows that each $A_{-\tau(\lambda)}$, $\lambda\in\dC_+$, is a maximal dissipative extension
of $A$ in $\gotH$. The family $\{A_{-\tau(\lambda)}\}_{\lambda\in\dC_+}$
is called the {\it \v{S}traus family of $A$ associated with $\tau$}
(cf. \cite{S70} and e.g. \cite[Section~3.3]{DHS03}) and for brevity we shall often call
$\{A_{-\tau(\gl)}\}$ simply {\it \v{S}traus family}.

Since $\cH$ is finite dimensional Fatous theorem (see \cite{Don,Gar}) implies that the limit
$\tau(\gl+i0)=\lim_{\epsilon\rightarrow+0}\tau(\gl+i\epsilon)$
from the upper half-plane exists for a.e. $\gl\in \bR$.
As in Section~\ref{scatsec} we denote set of real points $\lambda$ where this limit
exists by $\Sigma^\tau$. If there is no danger of confusion we will
usually write $\tau(\lambda)$ instead of $\tau(\lambda+i0)$ for $\lambda\in\Sigma^\tau$.
Obviously, the Lebesgue measure of $\bR \setminus \gS^\gt$ is zero.
Hence the \v{S}traus family
$\{A_{-\tau(\gl)}\}_{\gl\in\bC_+}$ admits a continuation to $\bC_+\cup
\Sigma^\tau$ which is also denoted by $\{A_{-\tau(\gl)}\}$, $\gl\in\bC_+\cup
\Sigma^\tau$. We remark that in the case $\imag(\tau(\lambda))=0$ for some $\lambda\in\dC_+\cup\Sigma^\tau$
the maximal dissipative operator $A_{-\tau(\gl)}$ is self-adjoint.

Let $M(\cdot)$ be the Weyl function of the boundary triplet
$\Pi=\{\cH,\Gamma_0,\Gamma_1\}$. Then $M(\cdot)$ is an $[\cH]$-valued Nevanlinna function
and $\imag(M(\lambda))$ is strictly positive for $\lambda\in \dC_+$.
Therefore
\begin{equation*}
N_{-\tau(\lambda)}(\gl) := -\bigl(\gt(\gl) + M(\gl)\bigr)^{-1}, \quad \gl \in \bC_+,
\end{equation*}
is a well-defined Nevanlinna function, see also \eqref{ntheta}.
The set of all real $\lambda$ where the limit
\begin{equation*}
N_{-\tau(\lambda+i0)}(\lambda+i0)
=\lim_{\epsilon\rightarrow +0} -\bigl(\gt(\gl+i\epsilon) + M(\gl+i\epsilon)\bigr)^{-1}
\end{equation*}
exists will for brevity be denoted by $\Sigma^N$.
Furthermore, for fixed $\lambda\in\Sigma^\tau$ we define an $[\cH]$-valued Nevanlinna function
$Q_{-\tau(\lambda)}(\cdot)$ by
\begin{equation}\label{qlam1}
Q_{-\tau(\gl)}(\mu) := -\bigl(\gt(\gl) + M(\mu)\bigr)^{-1},
\quad \mu \in \bC_+,
\end{equation}
and denote by $\Sigma^{Q_\lambda}$ the set of all real points $\mu$ where the limit
\begin{equation}\label{qlam2}
\quad Q_{-\tau(\gl)}(\mu+i0) =\lim_{\epsilon\rightarrow +0}Q_{-\tau(\lambda)}(\mu+i\epsilon)
\end{equation}
exists.
Notice that the complements $\bR \setminus \gS^{N}$ and $\bR \setminus
\gS^{Q_\lambda}$ are of Lebesgue measure zero. The next lemma will
be used in Section~\ref{mainsec}.
\begin{lem}\label{char}
Let $A$, $\Pi=\{\cH,\Gamma_0,\Gamma_1\}$, $M(\cdot)$ and $\tau(\cdot)$ be as above.
Then the following assertions {\rm (i)-(iii)} are true.
\begin{enumerate}

\item[{\em (i)}]
If $\gl \in \gS^\gt$ and $\mu \in \gS^M \cap \gS^{Q_\lambda}$, then the operator $\gt(\gl) + M(\mu)$
is invertible and
\begin{equation}\la{4.8}
\bigl(\gt(\gl) + M(\mu)\bigr)^{-1} =
\lim_{\epsilon\rightarrow +0}\bigl(\tau(\lambda)+M(\mu+i\epsilon)\bigr)^{-1}.
\end{equation}

\item[{\em (ii)}]
If $\gl \in \gS^\gt\cap\gS^M \cap \gS^{N}$, then the operator $\gt(\gl) + M(\gl)$
is invertible and
\begin{equation}\la{4.9}
\bigl(\tau(\lambda)+M(\lambda)\bigr)^{-1}=
\lim_{\epsilon\rightarrow +0}\bigl(\tau(\lambda+i\epsilon)+M(\lambda+i\epsilon)\bigr)^{-1}.
\end{equation}

\item[{\em (iii)}]
If $\gl \in \gS^\gt\cap\gS^M \cap \gS^N$, then $\gl \in \gS^{Q_\lambda}$ and
\begin{equation}\la{4.95}
\bigl(\tau(\lambda)+M(\lambda)\bigr)^{-1}=
\lim_{\epsilon\rightarrow +0}\bigl(\tau(\lambda)+M(\lambda+i\epsilon)\bigr)^{-1}.
\end{equation}
\end{enumerate}
\end{lem}
\begin{proof}
(i) If $\lambda\in\Sigma^\tau$, $\mu \in \gS^M$, then $\lim_{\epsilon\to+0}(\gt(\gl) + M(\mu + i\epsilon)) = \gt(\gl) + M(\mu)$.
Since
\begin{displaymath}
\bigl(\gt(\gl) + M(\mu+i\epsilon)\bigr)Q_{-\tau(\gl)}(\mu+i\epsilon) =
Q_{-\tau(\gl)}(\mu+i\epsilon) \bigl(\gt(\gl) + M(\mu+i\epsilon)\bigr)=
 -I_{\kH}
\end{displaymath}
for all $\epsilon>0$, we get
\begin{equation*}
-I_\kH = \bigl(\gt(\gl) + M(\mu)\bigr)Q_{-\tau(\gl)}(\mu) = Q_{-\tau(\gl)}(\mu)\bigl(\gt(\gl) + M(\mu)\bigr)
\end{equation*}
for $\gl \in \gS^\gt$ and $\mu \in \gS^M \cap \gS^{Q_\lambda}$ which
proves \eqref{4.8}.

(ii) For $\gl \in \gS^\gt \cap \gS^M$ clearly
\begin{equation*}
\lim_{\epsilon\to+0}\bigl(\gt(\gl + i\epsilon) + M(\gl + i\epsilon)\bigr) =
\gt(\gl) + M(\gl)
\end{equation*}
exists. Since
$(\gt(\gl) + M(\gl))N_{-\tau(\lambda)}(\gl) = N_{-\tau(\lambda)}(\gl)(\gt(\gl)
+ M(\gl))= -I_\kH$ for all $\gl \in \bC_+$ we have
\begin{equation*}
-I_\kH = \bigl(\gt(\gl) + M(\gl)\bigr)N_{-\tau(\lambda)}(\gl) = N_{-\tau(\lambda)}(\gl)
\bigl(\gt(\gl) + M(\gl)\bigr)
\end{equation*}
for  $\gl \in \gS^\gt \cap \gS^M \cap \gS^{N}$ which verifies
\eqref{4.9}.

(iii) Let $\gl \in \gS^\gt \cap \gS^M \cap \gS^{N}$. Let us show that
$\gl \in \gS^{Q_\lambda}$, i.e., we have to show that
$\lim_{\epsilon\to+0}(\gt(\gl) + M(\gl + i\epsilon))^{-1}$ exists.
Since $\gt(\gl) + M(\gl)$ is boundedly invertible and $\gt(\gl) +
M(\gl + i\epsilon)$, $\epsilon > 0$, converges in the operator norm
to $\gt(\gl) + M(\gl)$ the family
$\{(\gt(\gl) + M(\gl + i\epsilon))^{-1}\}_{\epsilon > 0}$ is uniformly bounded. Using
\begin{equation*}
\begin{split}
&\bigl(\gt(\gl) + M(\gl + i\epsilon)\bigr)^{-1} - \bigl(\gt(\gl) + M(\gl)\bigr)^{-1}\\
&\quad=
-\bigl(\gt(\gl) + M(\gl + i\epsilon)\bigr)^{-1}\bigl( M(\gl + i\epsilon) -
M(\gl)\bigr)\bigl(\gt(\gl) + M(\gl)\bigr)^{-1},
\quad \epsilon > 0,
\end{split}
\end{equation*}
one obtains the existence of $\lim_{\epsilon\to+0}(\gt(\gl) + M(\gl + i\epsilon))^{-1}$ and \eqref{4.95}.
\end{proof}
Let $A$, $\Pi=\{\cH,\Gamma_0,\Gamma_1\}$ and $M(\cdot)$ be as in the beginning of this
section and let as above $\tau(\cdot)$
be a matrix Nevanlinna function with values in $[\cH]$.
For each maximal dissipative operator from the \v{S}traus
family $\{A_{-\tau(\lambda)}\}_{\lambda\in\dC_+}$ the
characteristic function $W_{A_{-\tau(\lambda)}}$ is given by
\bea\label{2.17}
\lefteqn{
W_{A_{-\tau(\gl)}}:\dC_-\rightarrow [\cH_{\tau(\lambda)}]}\\
& &\mu\mapsto I_{\kH_{\tau(\gl)}}+2
iP_{\tau(\lambda)}\sqrt{\imag(\tau(\gl))}\bigl(\tau(\gl)^*+M(\mu)\bigr)^{-1}
\sqrt{\imag(\tau(\gl))}\upharpoonright_{\cH_{\tau(\lambda)}},
\nonumber
\eea
(see \cite{DM92} and \eqref{2.81}), where we have used
$\kH_{\tau(\gl)}=\ran(\imag(\gt(\gl)))$, $\lambda\in\Sigma^\tau$, and denoted the projection
and restriction onto $\cH_{\tau(\lambda)}$ by $P_{\tau(\lambda)}$ and $\upharpoonright_{\cH_{\tau(\lambda)}}$,
respectively.

If we regard the \v{S}traus family $\{A_{-\tau(\lambda)}\}$ on the larger set $\dC_+\cup\Sigma^\tau$,
then for $\lambda\in\Sigma^\tau$ the characteristic function $W_{A_{-\tau(\lambda)}}(\cdot)$
is defined as in \eqref{2.17}. Notice that in the case $\imag(\tau(\lambda))=0$ for $\lambda\in\Sigma^\tau$
the characteristic function of the self-adjoint extension $A_{-\tau(\lambda)}$ of $A$ is the identity
operator on the trivial space $\cH_{\tau(\lambda)}=\{0\}$.
Since the characteristic functions $W_{A_{-\tau(\lambda)}}(\cdot)$, $\gl \in
\dC_+\cup\gS^\gt$, are contractive $[\cH_{\tau(\lambda)}]$-valued
functions in the lower half-plane, the limits
\begin{equation*}
W_{A_{-\tau(\lambda)}}(\mu - i0)=\lim_{\epsilon\rightarrow +0}W_{A_{-\tau(\lambda)}}(\mu - i\epsilon)
\end{equation*}
exist for
a.e. $\mu \in \bR$, cf. \cite{FN}. The next proposition is a simple consequence of Lemma~\ref{char}.
\begin{prop}\label{charac}
Let $A$, $\Pi=\{\cH,\Gamma_0,\Gamma_1\}$ and $M(\cdot)$ be as above
and let $\tau(\cdot)$ be an $[\cH]$-valued Nevanlinna function. Let $\{A_{-\tau(\lambda)}\}_{\lambda\in\dC_+\cup\Sigma^\tau}$
be the \v{S}traus
family of maximal dissipative extensions of $A$ and let $W_{A_{-\tau(\lambda)}}(\cdot)$ be the corresponding
characteristic functions. Then the following holds.
\begin{itemize}
\item [{\rm (i)}] If $\gl \in \gS^\gt$ and $\mu \in  \gS^M \cap \gS^{Q_\lambda}$, then
the limit $W_{A_{-\tau(\lambda)}}(\mu - i0)$ exists and
\bea
\lefteqn{
W_{A_{-\tau(\lambda)}}(\mu - i0) = }\nonumber \\
& &I_{\kH_{\tau(\gl)}} +
2iP_{\tau(\lambda)}\sqrt{\imag(\tau(\gl))}(\gt(\gl)^* +
M(\mu)^*)^{-1}\sqrt{\imag(\tau(\gl))}\upharpoonright_{\cH_{\tau(\lambda)}}.
\nonumber
\eea
\item [{\rm (ii)}]
If $\gl \in \gS^\gt \cap \gS^M \cap \gS^N$, then the
limit $W_{A_{-\tau(\lambda)}}(\gl - i0)$ exists and
\bea
\lefteqn{
W_{A_{-\tau(\lambda)}}(\gl - i0) = }\nonumber \\
& &
I_{\kH_{\tau(\gl)}} +
2iP_{\tau(\lambda)}\sqrt{\imag(\tau(\gl))}(\gt(\gl)^* +
M(\gl)^*)^{-1}\sqrt{\imag(\tau(\gl))}\upharpoonright_{\cH_{\tau(\lambda)}}.
\nonumber
\eea
\end{itemize}
\end{prop}

\subsection{Coupling of symmetric operators and coupled scattering systems}\label{coupsection}

Let, as in the previous subsection $A$ be a densely defined closed simple symmetric
operator in $\gotH$ with equal finite deficiency indices and let $\Pi=\{\cH,\Gamma_0,\Gamma_1\}$
be a boundary triplet for $A^*$ with corresponding Weyl function $M(\cdot)$.
Let $\tau(\cdot)$ be an $[\cH]$-valued Nevanlinna function and assume in addition that $\tau$ can be realized as the
Weyl function corresponding to a densely defined closed simple symmetric operator $T$ in some
separable Hilbert space
$\gotG$ and a suitable boundary triplet $\Pi_T=\{\cH,\gY_0,\gY_1\}$ for $T^*$.
It is worth to note that the Nevanlinna function $\tau(\cdot)$ has this property if and only if $\imag(\tau(\lambda))$
is invertible for some (and hence for all) $\lambda\in\dC_+$ and

\begin{equation}\label{condrealw}
\lim_{y\rightarrow\infty}\frac{1}{y}\bigl(\tau(iy)h,h\bigr)=0\quad\text{and}\quad
\lim_{y\rightarrow\infty} y\,\imag\bigl(\tau(iy)h,h\bigr)=\infty
\end{equation}
hold for all $h\in\kH$, $h\not=0$, (see e.g. \cite[Corollary 2.5 and Corollary 2.6]{LT77} and \cite{DM91, Mal92}).

In the following the function $-\tau(\cdot)$ and the \v{S}traus family
\begin{equation}\label{straussfam}
A_{-\tau(\lambda)}=A^*\upharpoonright\ker\bigl(\Gamma_1+\tau(\lambda)\Gamma_0\bigr)
\end{equation}
are in a certain sense the counterparts of the dissipative matrix $D\in[\cH]$ and the corresponding
maximal dissipative extension $A_D$ from Section~\ref{sec3.1}. Similarly to Theorem~\ref{III.1} we
construct an "energy dependent dilation" in Theorem~\ref{coupling} below, that is, we find a self-adjoint operator
$\widetilde L$ such that
\begin{equation*}
P_\gotH\bigl(\widetilde L-\lambda)^{-1}\upharpoonright_{\gotH}=\bigl(A_{-\tau(\lambda)}-\lambda\bigr)^{-1}
\end{equation*}
holds.

First we fix a separable Hilbert space
$\gotG$,
a densely defined closed simple symmetric operator $T\in\cC(\gotG)$ and a boundary triplet
$\Pi_T=\{\cH,\gY_0,\gY_1\}$ for $T^*$ such that $\tau(\cdot)$ is the corresponding Weyl function.
We note that $T$ and $\gotG$
are unique up to unitary equivalence and the resolvent set $\rho(T_0)$ of the self-adjoint operator
$T_0:=T^*\upharpoonright\ker(\gY_0)$
coincides with the set $\mathfrak h(\tau)$ of points of holomorphy of $\tau$, cf. Section~\ref{weylsec}.
Since the deficiency indices of $T$ are $n_+(T)=n_-(T)=n$ it follows that
\begin{equation*}
L:=A\oplus T,\qquad \dom L=\dom A\oplus\dom T,
\end{equation*}
is a densely defined closed simple symmetric operator in the separable
Hilbert space $\gotL:=\gotH\oplus\gotG$
with deficiency indices $n_\pm(L)=n_\pm(A)+n_\pm(T)=2n$.

The following theorem has originally  been proved in \cite[$\S$ 5]{DHMS00}. For the sake
of completeness we present another proof that differs from the original one, cf. \cite{BL07}.

             \begin{thm}\label{coupling}
Let $A$, $\Pi=\{\cH,\Gamma_0,\Gamma_1\}$, $M(\cdot)$ and $\tau(\cdot)$ be as above,
let $T$ be a densely defined closed simple symmetric operator
in $\gotG$ and $\Pi_T=\{\cH,\gY_0,\gY_1\}$ be a boundary
triplet for $T^*$ with Weyl function $\tau(\cdot)$.
Then
\begin{equation}\label{athe}
\widetilde L=L^*\!\upharpoonright \left\{
f \oplus g \in\dom(L^*):
\begin{array}{l}
\Gamma_0 f-\gY_0 g = 0\\
\Gamma_1 f+\gY_1 g=0
\end{array}
\right\},
\end{equation}
is a self-adjoint operator in $\gotL$ such that
\begin{equation*}
P_\gotH\bigl(\widetilde L-\lambda)^{-1}\upharpoonright_{\gotH}=\bigl(A_{-\tau(\lambda)}-\lambda\bigr)^{-1}
\end{equation*}
holds for all $\lambda\in\rho(A_0)\cap\mathfrak h(\tau)\cap\mathfrak h(-(M+\tau)^{-1})$ and
the minimality condition
\begin{equation*}
\gotL =\clospa\bigl\{\bigl(\widetilde L-\lambda\bigr)^{-1}\gotH:
\lambda\in\bC\backslash\bR\bigr\}
\end{equation*}
is satisfied. Moreover, $\widetilde L$ is semibounded from below if and only if
$A_0$ is semibounded from below and $(-\infty,\eta)\subset\mathfrak h(\tau)$ for some $\eta\in\dR$.
\end{thm}

\begin{proof}
It is easy to see that $\widetilde\Pi=
\{\kH\oplus\kH,\widetilde\Gamma_0,\widetilde\Gamma_1\}$,
where $\widetilde\Gamma_0:=(\Gamma_0,\gY_0)^\top$ and
$\widetilde\Gamma_1:=(\Gamma_1,\gY_1)^\top$, is a boundary triplet
for $L^*=A^*\oplus T^*$. If $\gamma(\cdot)$ and $\nu(\cdot)$ denote the
$\gamma$-fields of $\Pi=\{\cH,\Gamma_0,\Gamma_1\}$ and $\Pi_T=\{\cH,\gY_0,\gY_1\}$,
respectively, then the $\gamma$-field
$\widetilde\gamma$ and Weyl function $\widetilde M$ of $\widetilde\Pi=
\{\kH\oplus\kH,\widetilde\Gamma_0,\widetilde\Gamma_1\}$ are given by
\begin{equation*}
\gl\mapsto\widetilde\gamma(\gl)=\begin{pmatrix}\gamma(\gl) & 0\\ 0 & \nu(\gl)
\end{pmatrix}
\quad\text{and}\quad\gl\mapsto\widetilde M(\gl)=\begin{pmatrix} M(\gl) & 0\\
0 & \tau(\gl)
\end{pmatrix},
\end{equation*}
$\lambda\in\rho(A_0)\cap\rho(T_0)$, $A_0=A^*\upharpoonright\ker(\Gamma_0)$,
$T_0=T^*\upharpoonright\ker(\gY_0)$.
A simple calculation shows that the relation
\begin{equation}\label{wttheta}
\Theta:=\left\{
\begin{pmatrix} (v,v)^\top\\ (w,-w)^\top
\end{pmatrix}:
v,w\in\kH\right\}\in\widetilde \kC(\kH\oplus\kH)
\end{equation}
is self-adjoint in $\kH\oplus\kH$, hence the operator
$L_{\Theta}=L^*\upharpoonright\widetilde\Gamma^{(-1)}\Theta$
is a self-adjoint extension of $L$ in $\gotL=\gotH\oplus\gotG$ and $L_\Theta$
coincides with $\widetilde L$ in \eqref{athe}. Hence, with
$L_0=L^*\upharpoonright\ker(\widetilde\Gamma_0)=A_0\oplus T_0$ we have
\begin{equation}\label{resof}
\bigl(\widetilde L-\gl\bigr)^{-1}=
(L_0-\gl)^{-1}+\widetilde\gamma(\gl)
\bigl(\Theta-\widetilde M(\gl)\bigr)^{-1}\widetilde\gamma(\overline\gl)^*,
\end{equation}
for all $\lambda\in\rho(\widetilde L)\cap\rho(L_0)$ by \eqref{2.8}.
Note that the difference of the resolvents of $\widetilde L$ and $L_0$
is a finite rank operator and therefore by well-known perturbation results
$\widetilde L$ is semibounded if and only if $L_0$ is semibounded, that is,
$A_0$ and $T_0$ are both semibounded. From $\rho(T_0)=\mathfrak h(\tau)$
we conclude that the last assertion of the theorem holds.

Similar considerations as
in the proof of Theorem~\ref{III.1} show that
\begin{equation}\label{thetam}
\bigl(\Theta-\widetilde M(\gl)\bigr)^{-1}=
-\begin{pmatrix} (M(\gl)+\tau(\gl))^{-1} & (M(\gl)+\tau(\gl))^{-1}\\
(M(\gl)+\tau(\gl))^{-1} & (M(\gl)+\tau(\gl))^{-1}
\end{pmatrix}
\end{equation}
holds for all $\lambda\in\rho(\widetilde L)\cap\rho(L_0)$. Therefore the
compressed resolvent of $\widetilde L$ has the form
\begin{equation*}
P_{\gotH}\bigl(L-\gl\bigr)^{-1}\upharpoonright\gotH
=(A_0-\gl)^{-1}-\gamma(\gl) \bigl(M(\gl)+\tau(\gl)\bigr)^{-1}\gamma(\overline\gl)^*
\end{equation*}
and coincides with $(A_{-\tau(\lambda)}-\lambda)^{-1}$ for all $\lambda$ belonging to
\begin{equation*}
\rho(L_0)\cap\rho(\widetilde L)=\rho(A_0)\cap\mathfrak h(\tau)\cap\mathfrak h\bigl(-(M+\tau)^{-1}\bigr),
\end{equation*}
see Section~\ref{weylsec}.
The minimality condition follows from
the fact that $T$ is simple, $\clospa\{\ker(T^*-\lambda):\lambda\in\dC\backslash\dR\}$, and
\eqref{resof} in a similar way as in the proof of Theorem~\ref{III.1}
\end{proof}

\begin{exam}
{\rm 
Let  $A$ be the symmetric Sturm-Liouville operator from Example~\ref{pavlov} and let 
$\Pi= \{\dC^n,\Gamma_0,\Gamma_1\}$ be the boundary triplet  for $A^*$  defined by
\eqref{3.40A}. Besides the operator $A$ we consider the minimal operator $T$ in $\gotG=L^2(\dR_-,\dC^n)$ 
associated to the Sturm-Liouville differential expression $-\tfrac{d^2}{dx^2}+Q_-$, 
\begin{equation*}
T=-\frac{d^2}{dx^2}+Q_-,\quad\dom (T)=\bigl\{g\in\cD_{\rm max,-}:g(0)=g^\prime(0)=0\bigr\}.
\end{equation*}
Analogously to Example~\ref{pavlov} it is assumed that $Q_-\in L^1_{\rm loc}(\dR_-,[\dC^n])$ satisfies 
$Q_-(\cdot)=Q_-(\cdot)^*$, that the limit point case prevails at $-\infty$ and the maximal domain 
$\cD_{\rm max,-}$ is 
defined in the same way as $\cD_{\max,+}$ in Example~\ref{pavlov} with $\dR_+$ and $Q_+$ replaced
by $\dR_-$ and $Q_-$, respectively. 

It is easy to see that $\Pi_T=\{\dC^n,\gY_0,\gY_1\}$, where 
\begin{equation}
\gY_0 g:=g(0),\quad \gY_1 g:=-g^\prime(0),\quad g\in\dom(T^*)=\cD_{\rm max,-},
\end{equation}
is a boundary triplet for $T^*$. For $f\in\dom (A^*)$ and $g\in\dom (T^*)$ the conditions 
$\Gamma_0f-\gY_0g=0$ and $\Gamma_1f+\gY_1g=0$ in \eqref{athe} stand for
\begin{equation*}
f(0+)=g(0-)\quad\text{and}\quad f^\prime(0+)=g^\prime(0-),
\end{equation*} 
so that the operator $\widetilde L$ in Theorem~\ref{coupling}
is the self-adjoint Sturm-Liouville operator
\begin{equation*}
\widetilde L=-\frac{d^2}{dx^2}+Q,\qquad Q(x)=
\begin{cases} Q_+(x), &  x\in\dR_+,\\
Q_-(x), & x\in\dR_-,
\end{cases}
\end{equation*}
in $L^2(\dR,\dC^n)$.
}
\end{exam}

Let $A$, $\Pi=\{\cH,\Gamma_0,\Gamma_1\}$, $M(\cdot)$ and $T$, $\Pi_T=\{\cH,\gY_0,\gY_1\}$,
$\tau(\cdot)$ be as in the beginning of this subsection. We define the
families $\{\cH_{M(\lambda)}\}_{\lambda\in\Sigma^M}$ and
$\{\cH_{\tau(\lambda)}\}_{\lambda\in\Sigma^\tau}$ of Hilbert spaces $\cH_{M(\lambda)}$
and $\cH_{\tau(\lambda)}$ by
\begin{equation}\la{4.2}
\cH_{M(\lambda)}=\ran\bigl(\imag(M(\lambda+i0))\bigr) \quad\text{and}\quad
\cH_{\tau(\lambda)}=\ran\bigl(\imag(\tau(\lambda+i0))\bigr)
\end{equation}
for all real points $\lambda$ belonging to $\Sigma^M$ and $\Sigma^\tau$,
respectively, cf. Section~\ref{scatsec}. As usual the projections and restrictions
in $\cH$ onto $\cH_{M(\lambda)}$ and $\cH_{\tau(\lambda)}$ are denoted by $P_{M(\lambda)}$,
$\upharpoonright_{\cH_{M(\lambda)}}$ and $P_{\tau(\lambda)}$,
$\upharpoonright_{\cH_{\tau(\lambda)}}$, respectively.

The next theorem is the counterpart of Theorem~\ref{dilscat} in the present framework.
We consider the complete scattering system $\{\widetilde L,L_0\}$
consisting of the self-adjoint operators $\widetilde L$ from Theorem~\ref{coupling}
and
\begin{equation*}
L_0:=A_0\oplus T_0, \quad A_0=A^*\upharpoonright\ker(\Gamma_0),\quad
T_0=T^*\upharpoonright\ker(\gY_0),
\end{equation*}
and express
the scattering matrix $\{\widetilde S(\lambda)\}$ in terms of the function $M(\cdot)$
and $\tau(\cdot)$.

\begin{thm}\label{scatcoup}
Let $A$, $\Pi=\{\cH,\Gamma_0,\Gamma_1\}$, $M(\cdot)$ and $T$, $\Pi_T=\{\cH,\gY_0,\gY_1\}$,
$\tau(\cdot)$ be as above. Define $\cH_{M(\lambda)}$, $\cH_{\tau(\lambda)}$ as in \eqref{4.2}
and let $L_0=A_0\oplus T_0$ and $\widetilde L$ be as in Theorem~{\rm \ref{coupling}}. Then the following holds.
\begin{itemize}
\item [{\rm (i)}] $L_0^{ac}=A^{ac}_0\oplus T_0^{ac}$ is unitarily equivalent to the multiplication operator
with the free variable in $L^2(\bR,d\gl,\kH_{M(\gl)}\oplus\cH_{\tau(\lambda)})$.
\item [{\rm (ii)}] In $L^2(\bR,d\gl,\kH_{M(\gl)}\oplus\cH_{\tau(\lambda)})$ the scattering
matrix $\{\widetilde S(\gl)\}$ of the complete scattering system $\{\widetilde L,L_0\}$ is given by
        \begin{equation} \label{4.13A}
\widetilde S(\gl) = I_{\kH_{M(\gl)}\oplus \kH_{\tau(\lambda)}}
-2i\begin{pmatrix} \widetilde T_{11}(\lambda) &  \widetilde T_{12}(\lambda)\\
 \widetilde T_{21}(\lambda) &  \widetilde T_{22}(\lambda)\end{pmatrix}\in[\cH_{M(\lambda)}\oplus\cH_{\tau(\lambda)}],
\end{equation}
for all $\gl \in\Sigma^M\cap\Sigma^{\tau}\cap\Sigma^N$, where
\begin{equation*}
\begin{split}
\widetilde T_{11}(\lambda)&=P_{M(\lambda)}\sqrt{\imag(M(\lambda))}
                 \bigl(M(\gl)+\tau(\lambda)\bigr)^{-1}\sqrt{\imag(M(\lambda))}\upharpoonright_{\cH_{M(\lambda)}},\\
\widetilde T_{12}(\lambda)&=P_{M(\lambda)}\sqrt{\imag(M(\lambda))}
                 \bigl(M(\gl)+\tau(\lambda)\bigr)^{-1}\sqrt{\imag(\tau(\lambda))}\upharpoonright_{\cH_{\tau(\lambda)}},\\
\widetilde T_{21}(\lambda)&=P_{\tau(\lambda)}\sqrt{\imag(\tau(\lambda))}\bigl(M(\gl)+\tau(\lambda)\bigr)^{-1}
          \sqrt{\imag(M(\lambda))}\upharpoonright_{\cH_{M(\lambda)}},\\
\widetilde T_{22}(\lambda)&=P_{\tau(\lambda)}\sqrt{\imag(\tau(\lambda))}\bigl(M(\gl)+\tau(\lambda)\bigr)^{-1}
\sqrt{\imag(\tau(\lambda))}
\upharpoonright_{\cH_{\tau(\lambda)}}\\
\end{split}
\end{equation*}
and $M(\lambda)=M(\lambda+i0)$, $\tau(\lambda)=\tau(\lambda+i0)$.
\end{itemize}
\end{thm}

\begin{proof}
Let $L=A\oplus T$ and let $\widetilde\Pi=\{\cH\oplus\cH,\widetilde\Gamma_0,\widetilde\Gamma_1\}$
be the boundary triplet for $L^*$ from the proof of Theorem~\ref{coupling}. The corresponding
Weyl function $\widetilde M$ is
\begin{equation}\label{wtweylm}
\lambda\mapsto\widetilde M(\lambda)=\begin{pmatrix} M(\lambda) & 0 \\ 0 & \tau(\lambda)
\end{pmatrix},\qquad\lambda\in\rho(A_0)\cap\rho(T_0),
\end{equation}
and since $L$ is a densely defined closed simple symmetric operator in the separable
Hilbert space $\gotL=\gotH\oplus\gotG$ we can apply Theorem~\ref{scattering}.
First of all we immediately conclude from
\begin{equation*}
\cH_{\widetilde M(\lambda)}=\cH_{M(\lambda)}\oplus\cH_{\tau(\lambda)},\qquad\lambda\in\Sigma^{\widetilde M}
=\Sigma^{M}\cap\Sigma^\tau,
\end{equation*}
that the absolutely continuous part
$L_0^{ac}=A_0^{ac}\oplus T_0^{ac}$ of $L_0$ is unitarily equivalent to the multiplication operator
with the free variable in the direct integral $L^2(\bR,d\gl,\kH_{M(\gl)}\oplus\cH_{\tau(\lambda)})$.
Moreover
\begin{equation}\label{slambda}
\widetilde S(\lambda)=I_{\widetilde\cH_\lambda}+2iP_{\widetilde M(\lambda)}\sqrt{\imag(\widetilde M(\lambda))}
\bigl(\Theta-\widetilde M(\lambda)\bigr)^{-1}\sqrt{\imag(\widetilde M(\lambda))}
\upharpoonright_{\cH_{\widetilde M(\lambda)}}
\end{equation}
holds for $\lambda\in\Sigma^{\widetilde M}\cap\Sigma^{N_\Theta}$,
where $\Theta$ is the self-adjoint relation from \eqref{wttheta}, the
set $\Sigma^{N_\Theta}$ is defined as in Section~\ref{scatsec} and
$P_{\widetilde M(\lambda)}$ and $\upharpoonright_{\cH_{\widetilde
M(\lambda)}}$ denote the projection and restriction in $\cH\oplus\cH$
onto $\cH_{\widetilde M(\lambda)}$, respectively.

For $\lambda\in\Sigma^{\widetilde M}\cap\Sigma^{N_\Theta}$ we have
\begin{equation*}
\lim_{\epsilon\rightarrow +0}\bigl(\Theta-\widetilde M(\lambda+i\epsilon)\bigr)^{-1}
=\bigl(\Theta-\widetilde M(\lambda+i0)\bigr)^{-1}
\end{equation*}
and
\begin{equation*}
\bigl(\Theta-\widetilde M(\gl)\bigr)^{-1}=
-\begin{pmatrix} (M(\gl)+\tau(\gl))^{-1} & (M(\gl)+\tau(\gl))^{-1}\\
(M(\gl)+\tau(\gl))^{-1} & (M(\gl)+\tau(\gl))^{-1}
\end{pmatrix},
\end{equation*}
cf. \eqref{thetam}. This implies that the sets $\Sigma^{\widetilde M}\cap\Sigma^{N_\Theta}$
and $\Sigma^M\cap\Sigma^\tau\cap\Sigma^N$ coincide. Moreover, by inserting the above
expression for $(\Theta-\widetilde M(\lambda))^{-1}$, $\lambda\in\Sigma^M\cap\Sigma^\tau\cap\Sigma^N$
into \eqref{slambda} and taking into account \eqref{wtweylm} we find that the scattering
matrix $\{\widetilde S(\lambda)\}$ of the scattering system $\{\widetilde L,L_0\}$ has
the form asserted in (ii).
\end{proof}

The following corollary, which is of similar type as Corollary~\ref{singlax}, is a simple
consequence of Theorem~\ref{scatcoup} and Proposition~\ref{charac}.

\begin{cor}\label{scatcoupcor}
Let the assumptions be as in Theorem~{\rm \ref{scatcoup}}, let $W_{A_{-\tau(\lambda)}}(\cdot)$ be
the characteristic function of the extension $A_{-\tau(\lambda)}$ in \eqref{straussfam} and assume in 
addition that $\sigma(A_0)$
is purely singular. Then $L_0^{ac}$ is unitarily equivalent to the multiplication operator
with the free variable in $L^2(\bR,d\gl,\cH_{\tau(\lambda)})$ and
the scattering matrix $\{\widetilde S(\gl)\}$ of the complete scattering system $\{\widetilde L,L_0\}$ is given by
\begin{equation*}
\begin{split}
\widetilde S(\lambda)&=W_{A_{-\tau(\lambda)}}(\lambda-i0)^*\\
&=I_{\kH_{\tau(\lambda)}}
-2iP_{\tau(\lambda)}\sqrt{\imag(\tau(\lambda))}\bigl(M(\gl)+\tau(\lambda)\bigr)^{-1}
\sqrt{\imag(\tau(\lambda))}
\upharpoonright_{\cH_{\tau(\lambda)}}
\end{split}
\end{equation*}
for a.e. $\lambda\in\dR$. In the special case $\sigma(A_0)=\sigma_p(A_0)$ this relation
holds for all $\lambda\in\Sigma^M\cap\Sigma^\tau\cap\Sigma^N$.
           \end{cor}

\begin{cor}
Let the assumptions be as in Corollary~{\rm\ref{scatcoupcor}} and suppose that the defect of $A$ is one, 
$n_{\pm}(A)=1$. 
Then
\begin{equation*}
\widetilde S(\lambda) = W_{A_{-\tau(\lambda)}}(\lambda-i0)^*
=\frac{M(\gl)+{\overline{\tau(\gl)}}}{M(\gl)+\tau(\gl)}
\end{equation*}
holds for a.e. $\lambda\in\dR$ with $\Imag\tau(\lambda+i0)\not=0$.
\end{cor}

\subsection{Scattering matrices of energy dependent and fixed dissipative scattering systems}\label{mainsec}

Let $A$, $\Pi=\{\cH,\Gamma_0,\Gamma_1\}$, $A_0=A^*\upharpoonright\ker(\Gamma_0)$
and $\tau(\cdot)$ be as in the previous subsections
and let $\{A_{-\tau(\lambda)}\}$ be the \v{S}traus family associated with $\tau$ from \eqref{straussfam}.
In the following we first fix some $\mu\in\dC_+\cup\Sigma^\tau$ and consider the fixed dissipative
scattering system $\{A_{-\tau(\mu)},A_0\}$. Notice that if $\mu\in\Sigma^\tau$ it may happen
that $A_{-\tau(\mu)}$ is self-adjoint. Let us denote by $\widetilde K_\mu$ the minimal
self-adjoint dilation of the maximal
dissipative extension $A_{-\tau(\mu)}$ in $\gotH\oplus L^2(\dR,d\lambda,\cH_{\tau(\mu)})$ constructed
in Theorem~\ref{III.1}. Here the fixed Hilbert space
$\cH_{\tau(\mu)}=\ran(\imag(\tau(\mu)))$ coincides with $\cH$ if $\mu\in\dC_+$
or $\cH_{\tau(\mu)}$ is a (possibly trivial) subspace of $\cH$ if $\mu\in\Sigma^\tau$.
Furthermore, if $K_0=A_0\oplus G_0$, where $G_0$ is the first order
differential operator in $L^2(\dR,d\lambda,\cH_{\tau(\mu)})$
from Lemma~\ref{disslem}, then
according to Theorem~\ref{dilscat} the absolutely continuous part $K_0^{ac}=A_0^{ac}\oplus G_0$ of $K_0$
is unitarily equivalent to the multiplication operator with the free variable in
the direct integral $L^2(\dR,d\lambda,\cH_{M(\lambda)}\oplus\cH_{\tau(\mu)})$ and
the scattering matrix $\{\widetilde S_\mu(\lambda)\}$ of the scattering system $\{\widetilde K_\mu,K_0\}$
is given by
\begin{equation}\label{smuscat}
\widetilde S_\mu(\lambda) = I_{\kH_{M(\lambda)}\oplus\kH_{\tau(\mu)}} -2i\begin{pmatrix}
\widetilde T_{11,\mu}(\lambda) &  \widetilde T_{12,\mu}(\lambda)\\
 \widetilde T_{21,\mu}(\lambda) &  \widetilde T_{22,\mu}(\lambda)\end{pmatrix}
\in\bigl[\cH_{M(\lambda)}\oplus\cH_{\tau(\mu)}\bigr],
\end{equation}
for all $\lambda \in\Sigma^M\cap\Sigma^{Q_\mu}$, where
\begin{equation*}
\begin{split}
\widetilde T_{11,\mu}(\lambda)&=P_{M(\lambda)}\sqrt{\imag(M(\lambda))}
                            \bigl(\tau(\mu) + M(\lambda)\bigr)^{-1}\sqrt{\imag(M(\lambda))}
\upharpoonright_{\cH_{M(\lambda)}},\\
\widetilde T_{12,\mu}(\lambda)&=P_{M(\lambda)}\sqrt{\imag(M(\lambda))}
                            \bigl(\tau(\mu)+ M(\lambda)\bigr)^{-1}\sqrt{\imag(\tau(\mu))}
\upharpoonright_{\cH_{\tau(\mu)}},\\
\widetilde T_{21,\mu}(\lambda)&=P_{\tau(\mu)}\sqrt{\imag(\tau(\mu))}\bigl(\tau(\mu)+ M(\lambda)\bigr)^{-1}
\sqrt{\imag(M(\lambda))}\upharpoonright_{\cH_{M(\lambda)}},\\
\widetilde T_{22,\mu}(\lambda)&=P_{\tau(\lambda)}\sqrt{\imag(\tau(\mu))}\bigl(\tau(\mu)+M(\lambda)\bigr)^{-1}
\sqrt{\imag(\tau(\mu))}\upharpoonright_{\cH_{\tau(\mu)}}\\
\end{split}
\end{equation*}
and $M(\lambda)=M(\lambda+i0)$. Here the set $\Sigma^{Q_\mu}$ and the corresponding
function $\lambda\mapsto Q_{-\tau(\mu)}(\lambda)$ defined in \eqref{qlam1}-\eqref{qlam2} replace
$\Sigma^{N_D}$ and $\lambda\mapsto (D-M(\lambda))^{-1}$ in
Theorem~\ref{dilscat}, respectively.

The following theorem is one of the main results of this paper. Roughly speaking it says that the
scattering matrix of the scattering system $\{\widetilde L,L_0\}$
from Theorem~\ref{scatcoup} pointwise coincides with scattering matrices of
scattering systems $\{\widetilde K_\mu,K_0\}$ of the above form.

\begin{thm}\label{main}
Let
$A$, $\Pi=\{\cH,\Gamma_0,\Gamma_1\}$, $M(\cdot)$ and $T$, $\Pi_T=\{\cH,\gY_0,\gY_1\}$,
$\tau(\cdot)$ be as in the beginning of Section~{\rm \ref{coupsection}} and let
$L_0=A_0\oplus T_0$ and $\widetilde L$ be as in Theorem~{\rm \ref{coupling}}.
For $\mu\in\Sigma^\tau$ denote the minimal self-adjoint dilation of $A_{-\tau(\mu)}$ in
$\gotH\oplus L^2(\dR,\cH_{\tau(\mu)})$ by $\widetilde K_\mu$ and let $K_0=A_0\oplus G_0$, where
$G_0$ is the self-adjoint first order differential operator in $L^2(\dR,\cH_{\tau(\mu)})$.

Then for each $\mu\in\Sigma^M\cap\Sigma^\tau\cap\Sigma^N$ the value of the scattering matrix
$\{\widetilde S_\mu(\lambda)\}$
of the scattering system $\{\widetilde K_\mu,K_0\}$ at energy $\lambda=\mu$ coincides with the value of the scattering matrix
$\{\widetilde S(\lambda)\}$
of the scattering system $\{\widetilde L,L_0\}$ at energy $\lambda=\mu$, that is,
\begin{equation}\label{klasse}
\widetilde S(\mu)=\widetilde S_\mu(\mu)\qquad \text{for all}\,\,\,\,\mu\in\Sigma^M\cap\Sigma^\tau\cap\Sigma^N.
\end{equation}
\end{thm}

\begin{proof}
According to Lemma~\ref{char}~(iii) each real $\mu\in\Sigma^M\cap\Sigma^\tau\cap\Sigma^N$
belongs also to the set $\Sigma^{Q_\mu}$. Therefore by comparing Theorem~\ref{scatcoup}
with the scattering matrix $\{\widetilde S_\mu(\lambda)\}$ of $\{\widetilde K_\mu,K_0\}$
at energy $\lambda=\mu$ in \eqref{smuscat} we conclude \eqref{klasse}.
\end{proof}

\begin{rem}
{\rm We note that Theorem~\ref{main} in a certain sense justifies the
use of self-adjoint dilations (or quasi-Hamiltonians) in the analysis
of scattering processes for open quantum systems. Indeed, if we
e.g. assume that the functions $M(\cdot)$, $\tau(\cdot)$ and
$(M(\cdot)+\tau(\cdot))^{-1}$ are continuous on an interval
$I\subset\dR$ containing the point $\mu$, then for $\lambda\in I$ the
scattering matrix $\{\widetilde S_\mu(\lambda)\}$ of the scattering
system $\{\widetilde K_\mu,K_0\}$ is a "good" approximation of the
"real" scattering matrix $\{\widetilde S(\lambda)\}$, $\lambda\in I$, of
the scattering system $\{\widetilde L,L_0\}$.
}
\end{rem}

\begin{rem}
{\rm
The statements of Theorem~\ref{scatcoup} and Theorem~\ref{main} are
also interesting from the viewpoint of inverse problems.
Namely, if $\tau(\cdot)$ is a matrix Nevanlinna function, satisfying
$\ker(\imag(\tau(\lambda)))=0$, $\lambda\in\dC_+$, and the conditions
\eqref{condrealw}, and if
$\{A_{-\tau(\lambda)},A_0\}$ is a family of energy dependent dissipative scattering
systems as considered above, then in general the Hilbert space $\gotG$ and the operators $T\subset T_0$
are not explicitely known, and hence also the scattering system $\{\widetilde L,L_0\}$ is not
explicitely known. However, according to Theorem~\ref{scatcoup} the scattering matrix $\{\widetilde S(\lambda)\}$
can be expressed in terms of $\tau(\cdot)$ and the Weyl function $M(\cdot)$, and by
Theorem~\ref{main} $\{\widetilde S(\lambda)\}$ can be obtained with
the help of the scattering matrices $\{\widetilde S_\mu(\lambda)\}$ of the scattering systems
$\{\widetilde K_\mu,K_0\}$.
}
\end{rem}

The following corollary concerns the scattering matrices $\{S_{-\tau(\mu)}(\lambda)\}$
of the energy dependent dissipative scattering systems
$\{A_{-\tau(\mu)},A_0\}$, $\mu\in\Sigma^\tau$.

\begin{cor}
Let the assumptions be as in Theorem~{\rm \ref{main}} and let $\mu\in\Sigma^M\cap\Sigma^\tau\cap\Sigma^N$.
Then the scattering matrix $\{S_{-\tau(\mu)}(\lambda)\}$ of the dissipative scattering
system $\{A_{-\tau(\mu)},A_0\}$ at energy $\lambda=\mu$ coincides with
the upper left corner of the scattering matrix $\{\widetilde S(\lambda)\}$ of the scattering system
$\{\widetilde L,L_0\}$ at energy $\lambda=\mu$.
\end{cor}

Let again $\widetilde K_\mu$ be the minimal self-adjoint dilation of the maximal dissipative
operator $A_{-\tau(\mu)}$ in $\gotH\oplus L^2(\dR,d\lambda,\cH_{\tau(\mu)})$.
In the next corollary we focus on the Lax-Phillips scattering matrices
$\{S^{LP}_\mu(\lambda)\}$ of the Lax-Phillips scattering systems $\{\widetilde K_\mu,\cD_{-,\mu},\cD_{+,\mu}\}$,
where
\begin{equation*}
\cD_{-,\mu}:=L^2\bigl(\dR_-,\cH_{\tau(\mu)}\bigr)\quad
\text{and}\quad \cD_{+,\mu}:=L^2\bigl(\dR_+,\cH_{\tau(\mu)}\bigr)
\end{equation*}
are incoming and outgoing subspaces for $\widetilde K_\mu$, cf. Lemma~\ref{III.3}.
If $W_{A_{-\tau(\mu)}}(\cdot)$ is the characteristic function of $A_{-\tau(\mu)}$, cf. \eqref{2.17}, then
according to Corollaries~\ref{lax1} and \ref{adamyanarov} we have
\begin{equation*}
\begin{split}
S^{LP}_\mu(\lambda)&=W_{A_{-\tau(\mu)}}(\lambda-i0)^*\\
&=I_{\cH_{\tau(\lambda)}}-2iP_{\tau(\lambda)}
\sqrt{\imag(\tau(\lambda))}
\bigl(\tau(\mu)+M(\lambda)\bigr)^{-1}\sqrt{\imag(\tau(\lambda))}\upharpoonright_{\cH_{\tau(\lambda)}}
\end{split}
\end{equation*}
for all $\lambda\in\Sigma^M\cap\Sigma^{Q_\mu}$, cf. Proposition~\ref{charac} and Corollary~\ref{scatcoupcor}.
Statements (ii) and (iii) of the following corollary can be regarded as generalizations of the classical
Adamyan-Arov result, cf. \cite{AA1,AA2,AA3,AA4} and Corollary~\ref{adamyanarov}.

\begin{cor}
Let the assumptions be as in Theorem~{\rm \ref{main}} and let
$\mu\in\Sigma^M\cap\Sigma^\tau\cap\Sigma^N$.
\begin{itemize}
\item [{\rm (i)}] The scattering matrix $\{S^{LP}_\mu(\lambda)\}$
of the Lax Phillips scattering
system $\{\widetilde K_\mu,\cD_{-,\mu},\cD_{+,\mu}\}$ at energy $\lambda=\mu$ coincides with
the lower right corner of the scattering matrix $\{\widetilde S(\lambda)\}$ of the scattering system
$\{\widetilde L,L_0\}$ at $\lambda=\mu$.
\item [{\rm (ii)}] The characteristic function
$W_{A_{-\tau(\mu)}}(\cdot)$ of $A_{-\tau(\mu)}$ satisfies
\begin{equation*}
\begin{split}
&S^{LP}_\mu(\mu)=W_{A_{-\tau(\mu)}}(\mu-i0)^*\\
&\,\,\,=I_{\cH_{\tau(\mu)}}-2iP_{\tau(\mu)}
\sqrt{\imag(\tau(\mu))}
\bigl(\tau(\mu)+M(\mu)\bigr)^{-1}\sqrt{\imag(\tau(\mu))}\upharpoonright_{\cH_{\tau(\mu)}}.
\end{split}
\end{equation*}
\item [{\rm (iii)}] If $\sigma(A_0)$ is purely singular, then
\begin{equation*}
\widetilde S(\mu)=S^{LP}_\mu(\mu)=W_{A_{-\tau(\mu)}}(\mu-i0)^*
\end{equation*}
holds for a.e. $\mu\in\dR$. In the special case $\sigma(A_0)=\sigma_p(A_0)$ this is true for
all $\mu\in\Sigma^M\cap\Sigma^\tau\cap\Sigma^N$.
\end{itemize}
\end{cor}

\subsection{A quantum transmitting Schr\"{o}dinger-Poisson system}\label{exam2}

As an example we consider an open quantum system of
similar type as in Section~\ref{III}. Instead of a single pseudo-Hamiltonian $A_D$ here
the open quantum system is described by a family of energy dependent pseudo-Hamiltonians
$\{A_{-\tau(\lambda)}\}$ which is sometimes called a quantum transmitting family.

Let, as in Section~\ref{III}, $(x_l,x_r)\subset\dR$ be a bounded interval and let
$A$ be the symmetric Sturm-Liouville operator
in $\gotH = L^2((x_l,x_r))$ given by
\begin{equation*}
\begin{split}
(Af)(x) &=
-\frac{1}{2}\frac{d}{dx}\frac{1}{m(x)}\frac{d}{dx}f(x) + V(x)f(x),\\
\dom(A) & =  \left\{f \in \gotH: \ba{l}
f, \frac{1}{m}f' \in W^1_2((x_l,x_r)) \\
f(x_l) = f(x_r)= 0\\
\left(\frac{1}{m}f'\right)(x_l) = \left(\frac{1}{m}f'\right)(x_r) = 0
\ea
\right\},
\end{split}
\end{equation*}
where $V,m, m^{-1}\in L^\infty((x_l,x_r))$ are real functions and $m>0$.
Let $v_l$, $v_r$ be real constants, let $m_l,m_r>0$ and define
$\widetilde V,\widetilde m\in L^\infty(\bR)$ by
\begin{equation}\label{vschl}
\widetilde V(x):=
\begin{cases}
v_l  & x \in (-\infty,x_l] \\
V(x) & x \in (x_l,x_r) \\
v_r  & x \in [x_r,\infty)
\end{cases}
\end{equation}
and
\begin{equation}\label{mschl}
\widetilde m(x):=
\begin{cases}
m_l & x \in (-\infty,x_l] \\
m(x) & x\in (x_l,x_r) \\
m_r & x\in [x_r,\infty)
\end{cases},
\end{equation}
respectively. We choose the boundary triplet $\Pi=\{\bC^2,\gG_0,\gG_1\}$,
\begin{equation*}
\gG_0f = \left(
\begin{array}{c}
f(x_l)\\
f(x_r)
\end{array}
\right),
\quad
\gG_1f = \left(
\begin{array}{c}
\left(\frac{1}{2m}f'\right)(x_l)\\
-\left(\frac{1}{2m}f'\right)(x_r)
\end{array}
\right),\quad f\in\dom(A^*),
\end{equation*}
from \eqref{2.13} for $A^*$.

In the following we consider the \v{S}traus family
\begin{equation*}
A_{-\tau(\lambda)}=A^*\upharpoonright\ker\bigl(\Gamma_1+\tau(\lambda)\Gamma_0\bigr),
\qquad\lambda\in\dC_+\cup\Sigma^\tau,
\end{equation*}
associated with the $2\times 2$-matrix Nevanlinna function
\begin{equation}\label{tauweyl}
\gl\mapsto\tau(\gl)=\left( \ba{cc}
i\sqrt{\frac{\gl - v_l}{2m_l}} & 0\\
0 & i\sqrt{\frac{\gl - v_r}{2m_r}} \ea \right);
\end{equation}
here the square root is defined on $\bC$ with a cut along $[0,\infty)$ and fixed by
$\imag(\sqrt{\gl})>0$ for $\lambda\not\in [0,\infty)$ and by $\sqrt{\lambda}\geq 0$ for
$\lambda\in[0,\infty)$, cf. Example~\ref{delta}, so that indeed $\imag(\tau(\lambda))>0$ for $\lambda\in\dC_+$
and $\overline{\tau(\lambda)}=\tau(\overline\lambda)$, $\lambda\in\dC\backslash\dR$.
Moreover it is not difficult to see that $\tau(\cdot)$ is holomorphic on
$\bC\backslash[\min\{v_l,v_r\},\infty)$ and $\gS^\gt = \bR$. The \v{S}traus family $\{A_{-\tau(\gl)}\}$, $\lambda\in\dC_+\cup\Sigma^\tau$,
has the explicit form
\begin{equation}\la{QT}
\begin{split}
\bigl(A_{-\tau(\gl)}f\bigr)(x) & :=
-\frac{1}{2}\frac{d}{dx}\frac{1}{m}\frac{d}{dx}f(x)+V(x)f(x), \\
\dom\bigl(A_{-\tau(\gl)}\bigr)&= \left\{f \in \gotH: \ba{l}
f, \frac{1}{m}f' \in W^1_2((x_l,x_r)),\\
\left(\frac{1}{2m}f'\right)(x_l) = -i\sqrt{\frac{\gl - v_l}{2m_l}}f(x_l),\\
\left(\frac{1}{2m}f'\right)(x_r) = i\sqrt{\frac{\gl - v_r}{2m_r}}f(x_r) \ea \right\}.
\end{split}
\end{equation}
The operator $A_{-\tau(\gl)}$ is self-adjoint if
$\gl \in (-\infty,\min\{v_l,v_r\}]$ and maximal dissipative if
$\gl\in (\min\{v_l,v_r\},\infty)$. We note that the
\v{S}traus family in \eqref{QT} plays an important role for the quantum transmitting
Schr\"odinger-Poisson system in \cite{klp} where it was
called the quantum transmitting family. For this open quantum system
the boundary conditions in \eqref{QT} are often called transparent boundary
conditions.

We leave it to the reader to verify that the Nevanlinna function $\tau(\cdot)$ in
\eqref{tauweyl} satisfies the conditions~\eqref{condrealw}. Hence by
\cite{DM91,LT77,Mal92} there exists a separable Hilbert space $\gotG$, a densely defined
closed simple symmetric operator $T$ in $\gotG$ and a boundary triplet
$\Pi_T=\{\dC^2,\gY_0,\gY_1\}$ for $T^*$ such that $\tau(\cdot)$ is the corresponding Weyl
function. Here $\gotG$, $T$ and $\Pi_T=\{\dC^2,\gY_0,\gY_1\}$ can be explicitly
described. Indeed, as Hilbert space $\gotG$ we choose
$L^2((-\infty,x_l)\cup(x_r,\infty))$ and frequently we identify this space with
$L^2((-\infty,x_l))\oplus L^2((x_r,\infty))$. An element $g\in\gotG$ will be written in
the form $g = g_l \oplus g_r$, where $g_l\in L^2((-\infty,x_l))$ and $g_r\in
L^2((x_r,\infty))$. The operator $T$ in $\gotG$ is defined by
\begin{equation*}
\begin{split}
(Tg)(x) := &
\left(
\ba{cc}
-\frac{1}{2}\frac{d}{dx}\frac{1}{m_l}\frac{d}{dx}g_l(x) + v_l g_l(x) & 0\\
0 & -\frac{1}{2}\frac{d}{dx}\frac{1}{m_r}\frac{d}{dx}g_r(x) + v_r g_r(x)
\ea
\right),\\
\dom(T) & := \left\{g=g_l\oplus g_r\in\gotG: \ba{l}
g \in W^2_2((-\infty,x_l))\oplus W^2_2((x_r,\infty))\\
g_l(x_l) = g_r(x_r) =
g'_l(x_l) = g'_r(x_l) = 0\\
\ea
\right\},
\end{split}
\end{equation*}
and it is well-known that $T$ is a densely defined closed simple
symmetric operator in $\gotG$ with deficiency indices
$n_+(T)=n_-(T)=2$. The adjoint operator $T^*$ is given by
\begin{equation*}
\begin{split}
(T^*g)(x) = &
\left(
\ba{cc}
-\frac{1}{2}\frac{d}{dx}\frac{1}{m_l}\frac{d}{dx}g_l(x) + v_l g_l(x) & 0\\
0 & -\frac{1}{2}\frac{d}{dx}\frac{1}{m_r}\frac{d}{dx}g_r(x) + v_r g_r(x)
\ea
\right),\\
\dom(T^*) & =\bigl\{g=g_l\oplus g_r\in\gotG: W^2_2((-\infty,x_l)) \oplus W^2_2((x_r,\infty))\bigr\}.
\end{split}
\end{equation*}
We leave it to the reader to check that $\Pi_T =\{\bC^2,\gY_0,\gY_1\}$, where
\begin{equation*}
\gY_0 g := \left(
\ba{c}
g_l(x_l)\\
g_r(x_r)
\ea
\right)
\quad \mbox{and} \quad
\gY_1 g := \left(
\ba{c}
-\frac{1}{2m_l}g'_l(x_l)\\
\frac{1}{2m_r}g'_r(x_r)
\ea
\right),
\end{equation*}
$g=g_l\oplus g_r\in \dom(T^*)$, is a boundary triplet for $T^*$. Notice that $T_0 =
T^*\!\upharpoonright\ker(\gY_0)$ is the restriction of $T^*$ to the domain
\begin{equation*}
\dom(T_0) = \bigl\{g\in\dom(T^*):
g_l(x_l) = g_r(x_r) = 0\bigr\},
\end{equation*}
that is, $T_0$ corresponds to Dirichlet boundary conditions. It is not difficult to see that
$\sigma(T_0)=[\min\{v_l,v_r\},\infty)$ and hence the Weyl function corresponding to
$\Pi_T =\{\bC^2,\gY_0,\gY_1\}$ is holomorphic on $\dC\backslash [\min\{v_l,v_r\},\infty)$.
\begin{lem}
Let $T\subset T^*$ and $\Pi_T=\{\bC^2,\gY_0,\gY_1\}$ be as above.
Then the corresponding Weyl function coincides with $\tau(\cdot)$
in \eqref{tauweyl}.
\end{lem}
\begin{proof}
A straightforward calculation shows that
\begin{equation*}
h_{l,\gl}(x) :=  \frac{i}{\sqrt{2m_l(\gl - v_l)}}
\exp\left\{-i\sqrt{2m_l(\gl - v_l)}(x -x_l)\right\}
\end{equation*}
belongs to $L^2((-\infty,x_l))$ for $\gl\in\bC\backslash[v_l,\infty)$ and satisfies
\begin{equation*}
-\frac{1}{2}\frac{d}{dx}\frac{1}{m_l}\frac{d}{dx}h_{l,\gl}(x)+v_lh_{l,\gl}(x)= \gl
h_{l,\gl}(x).
\end{equation*}
Analogously the function
\begin{equation*}
k_{r,\gl}(x) := \frac{i}{\sqrt{2m_l(\gl - v_r)}}
\exp\left\{i\sqrt{2m_r(\gl - v_r)}(x-x_r)\right\}
\end{equation*}
belongs to $L^2((x_r,\infty))$ for $\gl\in\bC\backslash[v_r,\infty)$ and satisfies
\begin{equation*}
-\frac{1}{2}\frac{d}{dx}\frac{1}{m_r}\frac{d}{dx}k_{r,\gl}(x)+v_rk_{r,\gl}(x)= \gl
k_{r,\gl}(x).
\end{equation*}
Therefore the functions
\begin{equation*}
h_\gl:=h_{l,\gl}\oplus 0\quad\text{and}\quad k_\gl:=0\oplus k_{r,\gl}
\end{equation*}
belong to $\gotG$ and we have $\ker(T^* -\gl)=\spann\{h_\gl,k_\gl\}$.

As the Weyl function $\widehat\tau(\cdot)$ corresponding to $T$ and $\Pi_T=\{\bC^2,\gY_0,\gY_1\}$
is defined by
\begin{equation*}
\gY_1 g_\gl=\widehat\gt(\gl)\gY_0 g_\gl\quad\text{for all}\quad g_\gl\in\ker(T^* -\gl),
\end{equation*}
$\gl\in\bC\backslash[\min\{v_l,v_r\},\infty)$, we conclude from
\begin{equation*}
\gY_1h_\gl = \frac{1}{2}\left( \ba{c}
-\frac{1}{m_l}\\
0 \ea \right) \quad \mbox{and} \quad \gY_0h_\gl = \left( \ba{c}
\frac{i}{\sqrt{2m_l(\gl - v_l)}}\\
0
\ea
\right)
\end{equation*}
and
\begin{equation*}
\gY_1k_\gl = \frac{1}{2}\left( \ba{c}
0\\
-\frac{1}{m_r}\\
\ea \right) \quad \mbox{and} \quad \gY_0k_\gl = \left( \ba{c}
0\\
\frac{i}{\sqrt{2m_r(\gl - v_r)}} \ea \right)
\end{equation*}
that $\widehat\tau$ has the form \eqref{tauweyl}, $\widehat\tau(\cdot)=\tau(\cdot)$.
\end{proof}

Let $A$, $\Pi=\{\bC^2,\Gamma_0,\Gamma_1\}$ and $T$, $\Pi_T=\{\dC^2,\gY_0,\gY_1\}$ be as above.
Then according to Theorem \ref{coupling} the operator
\begin{equation}\label{buslaev}
\widetilde L :=
A^*\oplus T^*\!\upharpoonright
\left\{f \oplus g \in \dom(A^*\oplus T^*):
\ba{l}
\Gamma_0 f- \gY_0 g = 0\\
\Gamma_1 f + \gY_1 g= 0
\ea
\right\}
\end{equation}
is a self-adjoint extension of $A\oplus T$ in $\gotH \oplus \gotG$. We can identify
$\gotH \oplus \gotG$ with
$L^2((-\infty,x_l))\oplus L^2((x_l,x_r)) \oplus L^2((x_r,\infty))$
and $L^2(\bR)$. The elements $f\oplus g$ in $\gotH \oplus \gotG$, $f\in\gotH$,
$g = g_l \oplus g_r \in \gotG$ will be written in the form
$g_l \oplus f \oplus g_r$.
The conditions $\Gamma_0 f=\gY_0 g$ and
$\Gamma_1 f=-\gY_1 g$, $f\in\dom(A^*)$, $g\in\dom(T^*)$,
have the form
\begin{equation*}
\left(
\ba{c}
f(x_l)\\
f(x_r)
\ea
\right) =
\left(
\ba{c}
g_l(x_l)\\
g_r(x_r)
\ea
\right)
\quad\text{and}\quad
\left(
\ba{c}
\left(\frac{1}{2m}f'_l\right)(x_l)\\
-\left(\frac{1}{2m}f'_r\right)(x_r)
\ea
\right) = \left(
\ba{c}
\frac{1}{2m_l}g'(x_l)\\
-\frac{1}{2m_r}g'(x_r)
\ea
\right).
\end{equation*}
Therefore an element $g_l\oplus f\oplus g_r$ in the domain of \eqref{buslaev}
has the properties
\begin{equation*}
g_l(x_l) = f(x_l)
\quad \mbox{and} \quad
f(x_r) = g_r(x_r)
\end{equation*}
as well as
\begin{equation*}
\frac{1}{m_l}g_l'(x_l) = \left(\frac{1}{m}f'\right)(x_l)
\quad \mbox{and} \quad
\left(\frac{1}{m}f'\right)(x_r) = \frac{1}{m_r}g_r'(x_r)
\end{equation*}
and the self-adjoint operator $\widetilde L$ in \eqref{buslaev} becomes
\bead
\lefteqn{
\widetilde L(g_l\oplus f\oplus g_r)=}\\
& &
\begin{pmatrix}
-\frac{1}{2}\frac{d}{dx}\frac{1}{m_l}\frac{d}{dx}g_l+v_lg_l & 0 & 0\\
0 & -\frac{1}{2}\frac{d}{dx} \frac{1}{m}\frac{d}{dx}f+Vf & 0 \\
0 & 0 &-\frac{1}{2}\frac{d}{dx}\frac{1}{m_r}\frac{d}{dx}g_r+v_rg_r
\end{pmatrix}.
\eead
With the help of \eqref{vschl} and \eqref{mschl} we see that \eqref{buslaev}
can be regarded as the usual self-adjoint second order differential operator
\begin{equation*}
\widetilde L = -\frac{1}{2}\frac{d}{dx}\frac{1}{\widetilde m}\frac{d}{dx}+ \widetilde V
\end{equation*}
on the maximal domain in $L^2(\bR)$, that is, \eqref{buslaev} coincides with the so-called
Buslaev-Fomin operator from \cite{klp}.

Denote by $M(\cdot)$ the Weyl function corresponding to $A$ and the boundary triplet $\Pi=\{\dC^2,\Gamma_0,\Gamma_1\}$,
cf. \eqref{fundsol}-\eqref{fsm}.
Since $\sigma(A_0)$ consists of eigenvalues Corollary~\ref{scatcoupcor} implies that the scattering matrix
$\{\widetilde S(\lambda)\}$ of the scattering system $\{\widetilde L,L_0\}$, $L_0=A_0\oplus T_0$, is given by
\begin{equation*}
\widetilde S(\lambda)=I_{\cH_{\tau(\lambda)}}-2i
P_{\tau(\lambda)}\sqrt{\imag(\tau(\lambda))}\bigl(M(\gl)+\tau(\lambda)\bigr)^{-1}
\sqrt{\imag(\tau(\lambda))}
\upharpoonright_{\cH_{\tau(\lambda)}}
\end{equation*}
for all $\lambda\in\rho(A_0)\cap\Sigma^N$, where
\begin{equation*}
\cH_{\tau(\lambda)}=\ran(\imag(\tau(\lambda)))=\begin{cases}
\{0\}, & \lambda\in(-\infty,\min\{v_l,v_r\}],\\
\dC, & \lambda\in (\min\{v_l,v_r\},\max\{v_l,v_r\}],\\
\dC^2, &  \lambda\in (\max\{v_l,v_r\},\infty).
\end{cases}
\end{equation*}
The scattering system $\{\widetilde L,L_0\}$ was already investigated in \cite{baro1,klp}.
There it was in particular shown that the scattering matrix $\{\widetilde S(\lambda)\}$ and
the characteristic function $W_{A_{-\tau(\gl)}}(\cdot)$ of the maximal dissipative extension
$A_{-\tau(\gl)}$ from \eqref{QT} are connected via
\begin{equation*}
\widetilde S(\lambda)=W_{A_{-\tau(\lambda)}}(\lambda-i0)^*,
\end{equation*}
which we here immediately obtain from Corollary~\ref{scatcoupcor}.

\end{document}